\documentclass[twocolumn,smallcondensed,smallextended,dvips]{svjour3}
\usepackage{mathptmx}

\usepackage{graphics}
\long\def\beginpgfgraphicnamed#1#2\endpgfgraphicnamed{\includegraphics{#1}}
\usepackage{microtype}
\usepackage{fixltx2e}

\usepackage[centertags,intlimits]{amsmath}
\usepackage{amssymb,amsfonts}
\usepackage{mathtools}

\DeclareMathOperator{\e}{e}
\renewcommand{\d}{\operatorname{d}\!}

\newcommand{\order}{\ensuremath{\mathcal{O}}}
\newcommand{\const}{\mathrm{const.}}
\newcommand{\abs}[1]{\left\lvert#1\right\rvert}

\journalname{Journal of Statistical Physics}

\begin{document}

\title{\large\bfseries Diffusion-limited reactions and mortal random
  walkers in confined geometries}
\author{Ingo Lohmar \and Joachim Krug}
\institute{I. Lohmar \and
           J. Krug \at
           Institute for Theoretical Physics, University of Cologne, Germany \\
           Tel.: +49-221-470-2818\\
           Fax:  +49-221-470-5159\\
           \email{il@thp.uni-koeln.de, krug@thp.uni-koeln.de}}
\date{Received: date / Accepted: date}
\maketitle

\begin{abstract}
  Motivated by the diffusion-reaction kinetics on interstellar dust
  grains, we study a first-passage problem of mortal random walkers in
  a confined two-dimensional geometry.  We provide an exact expression
  for the encounter probability of two walkers, which is evaluated in
  limiting cases and checked against extensive kinetic Monte Carlo
  simulations.  We analyze the continuum limit which is approached
  very slowly, with corrections that vanish logarithmically with the
  lattice size.  We then examine the influence of the shape of the
  lattice on the first-passage probability, where we focus on the
  aspect ratio dependence: Distorting the lattice always reduces the
  encounter probability of two walkers and can exhibit a crossover to
  the behavior of a genuinely one-dimensional random walk.  The nature
  of this transition is also explained qualitatively.
\end{abstract}

\section{Introduction}

Random walks are ubiquitous in statistical physics as well as in the
description of many natural phenomena.
One key example is the study of diffusion-reaction systems.

In our particular case, one motivation for the following examination
comes from a long-standing astrophysical puzzle, namely why there can
be a large abundance of molecular hydrogen in interstellar clouds of
gas and dust compared to the concentration of atomic hydrogen.  For
realistic conditions, the generally accepted mechanism is as follows
\cite{Gould63,Hollenbach70,Hollenbach71,Hollenbach71a}: Reactants
impinge onto the surface of dust grains in a homogeneous fashion.
They diffuse on the surface.  Additionally, they desorb at a certain
rate.  If two such atoms meet during their traversal of the surface,
they react to form a hydrogen molecule which immediately desorbs.  We
will sometimes refer to the \emph{recombination efficiency} of such a
system, defined as the fraction of impinging atoms that eventually
react to yield a molecule.

To study the catalytic role of the dust grains in the astrophysically
relevant regime of a small average number of atoms on the grain, one
sets up a zero-dimensional master equation for the distribution of the
number of reactants on a single grain, including terms for deposition,
desorption and reaction, and solves for the stationary state of the
system \cite{Green01,Biham02}.  Rate equations for the \emph{mean}
adatom number do not suffice as they cannot account for crucial
fluctuations and the non-\textsc{Poisson}ian statistics of the
reactant number that is induced by their confinement to a finite
closed surface on which the reaction takes place
\cite{Biham02,Lohmar06}.  Within such a master equation treatment,
apart from obvious single-atom rates, there only occurs a single
two-particle parameter, viz.\ the \emph{sweeping rate} describing how
often a pair of atoms on the grain meets.  One convenient way to
express this parameter is by means of the \emph{encounter probability}
originally introduced in \cite{Krug03} in a continuum setting.  This
is the key quantity of a \emph{first-passage problem} of two
\emph{mortal} random walkers: Mortality (i.e., terminating a random
walk with a certain probability per step) corresponds to desorption of
the atoms on the surface, the random walk to their diffusion, and the
question (of first-passage nature) is whether the two walkers meet
before either one dies, quantified by the encounter probability.

In general, a realistic full diffusion-reaction model has to be
examined by microscopic Monte Carlo simulations (see
e.g.~\cite{Chang05}).  This is mostly owed to the fact that
applications, and especially the astrophysical problem we described,
call for the inclusion of disorder in the local rates of hopping and
desorption, a feature hard to tackle analytically in more than one
dimension.  Despite the vast literature on diffusion properties in
disordered media this even holds true (to the best of our knowledge)
when we focus on the basic level of the encounter probability.

Still, it is useful to start with the analysis of simple homogeneous
models that are accessible rather easily, both to establish a
reference point for the validity of simulations and to make progress
in the analytical exploration.  On the level of the encounter
probability this means we can employ results from the theory of
homogeneous random walks.  Furthermore, the discrete random walk model
allows us to extend the generality step by step to different lattice
types and shapes, hopefully providing deeper insight into the
importance of these factors.  Finally, it is expected that the problem
will have some fundamental appeal for the theory of random walks in
relation to reaction-diffusion systems in general.

Examples from seemingly remote fields to which this analysis might
bear relevance include second-layer nucleation in epitaxial crystal
growth \cite{Krug03}, chemical kinetics inside aerosol droplets
\cite{Lushnikov03}, biophysical problems like exciton trapping on
photosynthetic units \cite{Montroll69} and a range of search,
transport and binding processes of or along DNA strands
\cite{Slutsky03,Slutsky04}.  Admittedly, in the latter case the
homogeneous situation is only a first step to meaningful results, and
the crucial two-dimensionality of our problem does not necessarily
apply either.

Our goal in this paper is thus to further elucidate the meaning of the
encounter probability, examining the pitfalls of the continuum limit
to aid comparison to Monte Carlo simulations, and to analyze the
influence of the lattice type and, most importantly, different
geometries.  We strive for a completely analytic theory to fully
explain all findings of simulations.  To this end, we will proceed as
follows: Section~\ref{sec:rw} will first introduce the model and basic
definitions along with our notation.  We will then show how the
fundamental quantity that we call `encounter probability' can be
obtained from a simple random walk calculation, and will analyze its
asymptotic behavior.  In Section~\ref{sec:comparisons} we investigate
the subtle continuum limit which implies a logarithmically slow
convergence to results obtained in a continuum model, which is crucial
for comparison to simulations.  We supplement this by simulation
results in Section~\ref{sec:simulations}.  Section~\ref{sec:shape} we
extend the random walk calculations to a rectangular lattice to
examine the role of the lattice shape, analyzing a
quasi-one-dimensional limit and deriving and explaining the effect of
a distorted aspect ratio.  Finally, we present our conclusions.

\section{Random walk treatment}
\label{sec:rw}

\subsection{Model and definitions}

We consider a homogeneous two-dimensional lattice with periodic
boundary conditions, meaning that all $S=L_1\cdot L_2$ sites are
equivalent.  Two walkers are randomly deposited on two (not
necessarily different) sites of the lattice, then they move between
nearest-neighbor sites with an undirected hopping rate $a$, and may
desorb at a constant rate $W\ll a$.  Whenever the two meet by one
hopping onto the site occupied by the other (or by coinciding initial
positions), they react and this random walk realization has ended
successfully; the probability (averaged over random walk realizations
and initial positions) at which this happens before either of the two
walkers desorbs is called the \emph{encounter probability} $p$
(subscripts will denote certain models and\,/\,or simulations from
which it is obtained, as well as further specializations).

Here we will treat a discrete-time random walk, although a
continuous-time version (CTRW) with an appropriate waiting time
distribution might seem closer to the physical system at hand.  We do
this in order to keep the analytical treatment as simple as possible,
and we will later argue and numerically prove
(Section~\ref{sec:simulations}) that the discrete version accurately
describes our situation.  Due to homogeneity of the lattice, the
problem in fact reduces to that of a single walker meeting a target
site, cf.\ Appendix~\ref{app:prw}.

\subsection{Exact results}

Quite generally, we are concerned with finite lattices translationally
invariant in all directions, which extend to $L_j$ lattice sites in
the $j$th of $d$ dimensions before periodically continuing.  The exact
expression of the random-walk encounter probability on such a lattice
is re-derived in Appendix~\ref{app:prw} and reads
\cite{denHollander82}
\begin{equation}
  \label{general-prw}
  \begin{split}
    p_\mathrm{rw}^{-1} &= (1-\xi) S P^*(\vec 0;\xi) \\
    &=\sum_{\vec m\in\Omega}
    \frac{1-\xi}{1-\xi\lambda(2\pi\mathbf L^{-1}\vec m)}.
  \end{split}  
\end{equation}
Here,
\begin{equation}
  \xi = 1 - \frac{W}{a+W} = \frac{a}{a+W} = \frac{1}{1+W/a}\lesssim 1
\end{equation}
is the survival probability per step, $S=\prod_{j=1}^d L_j$ is the
total number of sites, and $P^*(\vec 0;\xi)$ is the number of times a
mortal walker on a periodic homogeneous lattice returns to the origin.
$\Omega$ denotes the lattice, $\vec m$ is a `lattice vector' of $d$
integer components $0\leq m_j\leq L_j-1$ with $j=1,\dots,d$, and
$\mathbf L= \operatorname{diag}(L_1,\dots,L_d)$.  Finally, $\lambda$
is the \emph{structure function} (basically a discrete
\textsc{Fourier} transform of the normalized transition probability)
of the walk.  We specify to $d=2$ at this point, where it reads
$\lambda(\vec k)=(\cos k_1+\cos k_2)/2$ for an isotropic walk on a
square lattice that we will further on label as being of `type (a)',
and $\lambda(\vec k)=[\cos k_1+\cos k_2+\cos(k_1+k_2)]/3$ for the
isotropic walk on a triangular lattice (coordination number 6), now
designated as `type (b)'.

We focus on the two-dimensional case with both lattice lengths much
larger than unity and with `long survival' defined by $1-\xi\ll1$.
The expression~\eqref{general-prw} then affords several regimes,
characterized by the comparison of dimensionless `lengths'.
Introducing the typical single-atom random walk length
\begin{equation}
  \ell =\sqrt{a/W} = \sqrt{\frac{\xi}{1-\xi}}\gg1
\end{equation}
and with lattice dimensions $L_{1,2}\gg1$, one can associate $1\ll\ell
\ll L_{1,2}$ with `large' lattices, and $1\ll L_{1,2}\ll\ell $ with
`small' lattices.  The intermediate regime in which one lattice length
is smaller, yet the other larger than the random walk length is to be
discussed in a later Section.

For a still fairly large class of walks that includes the two cases
(a) and (b), one summation in equation~\eqref{general-prw} can be
carried out explicitly \cite{Montroll69}.  Appendix~\ref{app:one-sum}
gives the result, which we generalized to include the case $L_1\neq
L_2$.  Whenever we need to numerically evaluate $p_\mathrm{rw}$ we use
the resulting single-sum expression, further simplified for the two
lattice types, and implemented in a small \texttt{GNU Octave} script.

The encounter probability as given above allows for the two random
walkers to start on one site, and this counts as an encounter on the
zeroth step.  For applications and comparison to other models or
simulations, we will often need to deal with the encounter probability
$\tilde p_\mathrm{rw}$ calculated such that it only accounts for
meeting of the walkers by hopping, and that does \emph{not} allow the
initial condition of walkers starting at the same site.  The two
quantities are related by
\begin{equation}
  p_\mathrm{rw}= \frac1S +\left(1-\frac1S\right) \tilde p_\mathrm{rw},
\end{equation}
as becomes obvious by splitting up $p_\mathrm{rw}$ according to the
starting site of the second walker, i.e., either on the same site as
the first one (with probability $1/S$), or on any other site (with
probability $1-1/S$).  This leads to the expression
\begin{equation}
  \tilde p_\mathrm{rw} = \frac{S p_\mathrm{rw}-1}{S-1}
\end{equation}
in terms of $p_\mathrm{rw}$, and $\tilde p$ with other subscripts will
henceforth denote probabilities $p$ that are obtained from other
models using the corresponding analogous convention, i.e., excluding
an encounter due to coinciding initial positions.  In the context of
hydrogen recombination on dust grain surfaces, the corresponding
mechanism is the `Langmuir-Hinshelwood rejection' of atoms that
impinge on top of another; see e.\,g.\ \cite{Langmuir18} and
references therein for early original work.

\subsection{Large lattice approximation}\label{sec:prw-large-lattice}
Large lattices, formally given by $S\gg a/W\gg1$ or, more precisely,
by $1\ll\ell \ll L_{1,2}$, can equally well be defined as the regime
in which boundary conditions, and (apart from the initial placement)
the overall number of sites of the lattice no longer matter at all --
even if the boundaries truly affected the random walker (say, by
reflection), it \emph{either} approaches the target \emph{or}
experiences the finiteness and (possibly) boundedness of the lattice,
but never during one single walk.  It is thus safe to simply send
$L_{1,2}\to\infty$ in $P^*(\vec 0;\xi)$ in~\eqref{general-prw}, by
which procedure the sum becomes an area integral ($P(\vec 0;\xi)$ in
the standard notation used in the Appendices).  Some more details on
the ensuing approximations are given in
Appendix~\ref{app:large-grain-prw}, with the result that
\begin{equation}
p_\mathrm{rw}\approx \frac{a}{SW}
\begin{cases}
\pi \frac{1}{\ln[8a/W]} & \text{square lattice,}\\
\frac{2\pi}{\sqrt{3}} \frac{1}{\ln[12a/W]} & \text{triangular lattice,}
\end{cases}
\end{equation}
with relative errors of $\order(1-\xi)$.

\subsection{Small lattice approximation}\label{sec:rw-approx}
Here we have $1\ll L_{1,2}\ll\ell $ or $a/W\gg S\gg1$.
Again, we employ results for $P^*(\vec 0;\xi)$ from the random walk
literature.

For the moment, we restrict ourselves to the case $L_1=L_2=\sqrt{S}$.
The expansion for $1-\xi\ll S^{-1}\ll1$ then reads
\begin{equation}\begin{split}
    P^*(\vec 0;\xi) &=\frac1{S(1-\xi)} +c_1\ln(cS) \\
    &\quad +\order\left(S^{-1},S(1-\xi),\sqrt{1-\xi})\right)\,,
\end{split}\end{equation}
where $c$ and $c_1$ are real constants.  This is shown in
\cite{Montroll69}, also presenting the first calculations that deliver
upon the important pre-factor inside the logarithm (for square and
triangular lattices), which were extended and subject to minor
corrections in \cite{denHollander82} (cf.\ the earlier and easily
accessible derivation in \cite{Montroll65}, which unfortunately does
not give this pre-factor).  For the encounter probability one thus
obtains
\begin{equation}\label{small-lattice-prw}
p_\mathrm{rw}\approx 1-\frac{SW}{a}c_1\ln(cS)\,.
\end{equation}
It is important to note that the original expansion is valid in the
regime $(1-\xi)S\pi^{-1}\ln(cS)\ll1$, or equivalently,
$1-p_\mathrm{rw}\ll1$.  This would thus be a more precise definition
of a `small lattice'.

The constant $c_1$ has the value $1/\pi$ for the square lattice case
(a), and $c_1=\sqrt{3}/(2\pi)$ for the triangular lattice (b),
respectively.  The crucial factor $c$ \emph{inside} the logarithm
appears in the different guise $c_2$ in \cite{Montroll69}, related to
ours by $c_2/c_1=\ln c$.  For the square lattice the ratio yields
\begin{equation}\begin{split}
\ln c &= \frac{\pi}{3}+2(\gamma-\ln\pi+\frac12\ln 2) \\
&\quad
+4\left[e^{-2\pi}+\frac32e^{-4\pi}+\frac43e^{-6\pi}+\dots\right]\\
&\approx 0.612807020\,,  
\end{split}
\end{equation}
while for the triangular lattice this becomes
\begin{equation}\begin{split}
\ln c &= \frac{\pi}{2\sqrt{3}}+2(\gamma-\ln\pi+\frac12\ln 3) \\
&\quad-4\left[e^{-\sqrt{3}\pi}-\frac32e^{-2\sqrt{3}\pi}+\frac43e^{-3\sqrt{3}\pi}
  -\dots\right]\\
&\approx 0.853262084\,,
\end{split}
\end{equation}
where we used standard rounding and checked the consistency of the
numerical evaluation against the original figures and, when available,
improved ones from \cite{denHollander82}.  Here and in what follows,
$\gamma\approx0.5772156649$ denotes \textsc{Euler}'s constant.

\section{Continuum limit}
\label{sec:comparisons}

A natural question to ask is whether the model described above yields
a reasonable continuum limit for $L_{1,2}\to\infty$; on the one hand,
since such continuum limits are of genuine theoretical interest in
themselves, on the other hand, since we have earlier solved a
continuum model \cite{Lohmar06} to compare against.  The answer will
also provide important information for the comparison between
analytical theory and simulations.

The proper scaling of the discrete random walk parameters then
proceeds as follows.  We want to stay in the same regime of the
system, which means we need to keep both ratios $L_j/\ell $, and
consequently $SW/a$, fixed.  Further, we do not want to distort the
given aspect ratio of the lattice, so that $L_1/L_2$ is kept constant
as well (a moot point as long as we use quadratic lattices, i.e.,
$L_1\equiv L_2$ anyhow).  With these constraints we let the number of
adsorption sites become very large, $S\to\infty$.  Such a joint limit
preserves all quantities in our expressions that only depend on the
`regime parameter' $\sim S(1-\xi)$.  The limiting behavior can thence
be determined from the corresponding approximate results for the two
regimes separately.

In the continuum (or ``diffusion'') model proposed in \cite{Lohmar06},
a stationary diffusion equation for the probability density of a
moving atom is solved on a spherical surface, and given an absorbing
`target area'.  The encounter probability is then calculated as the
(suitably normalized) diffusion current entering this area.  The
results will be denoted by $p_\mathrm{diff}$.\footnote{Note that the
  cited article differs in that it uses this notation for only a
  certain contribution to the full encounter probability.}

We will also refer to results for $p$ obtained from random walk
theory, however in a heuristic fashion \cite{Lohmar06}, using known
expressions for $N_\mathrm{dis}$, the average number of distinct sites
visited by a random walker after it has taken a given number of steps.
The asymptotic result is that after $n\gg1$ steps (on a
two-dimensional regular lattice),
\begin{equation}
  N_\mathrm{dis}\approx\frac{\pi n}{C \ln(B n)}
\end{equation}
with real positive constants $C$ and $B$ depending on the lattice
type, viz.\ $C=1$ and $B=8$ for the square, and $C=\sqrt{3}/{2}$ and
$B=12$ for a triangular lattice, respectively (see e.g.\
\cite{Hughes95}).  The encounter probability derived on this route
will be labeled $p_\mathrm{rw,heur}$.  Note that the above
$N_\mathrm{dis}$ is the leading term of an asymptotic expansion, and
thence a poor estimate for moderate values of $n$.

In the comparison of analytical results, we allow for walkers to meet
by coinciding initial positions throughout, and we incorporate
corresponding terms in the diffusion models as well; in short, we
always use $p$ and not $\tilde p$.

\subsection{Large lattices}
\label{sec:large-grains}

The random walk result $p_\mathrm{rw}$ has been given above.  Its
diffusion model analogue reads
\begin{equation}\label{pdiff-large-sphere}
p_\mathrm{diff}\approx\frac{4ga}{SW}\frac1{\ln(4a/W)-2\gamma}\,,
\end{equation}
where $g$ denotes a lattice-dependent factor; it is $g=\pi/4$ for the
square lattice, and $g=\pi/(2\sqrt{3})$ for the triangular lattice,
respectively.  The heuristic result obtained from $N_\mathrm{dis}$
is
\begin{equation}
  p_\mathrm{rw,heur}\approx \frac{\pi a}{CSW}\frac{1}{\ln(Ba/W)}\,,
\end{equation}
with $B$, $C$ as defined above.  Since comparison between the factors
$g$ and $C$ immediately shows that for our cases $4g=\pi/C$, we see
that not only the functional dependence, but also the numerical
pre-factors of all three expressions $p_\mathrm{rw}$,
$p_\mathrm{diff}$, and $p_\mathrm{rw,heur}$ coincide.

As far as the factor \emph{inside} the logarithm is concerned, it is
not too surprising that $p_\mathrm{rw} $ and $p_\mathrm{rw,heur}$
agree as well.  However, the diffusion model expression
$p_\mathrm{diff}$ clearly differs here, though this becomes irrelevant
in the true limit $S\to\infty$.  Note that there is no longer any
imprint of the lattice type, as opposed to the discrete result.  We
should emphasize that in~\cite{Lohmar06} care was taken that in the
given form of the asymptotic expressions, all omitted terms are of
higher order, particularly in the denominator expression, where
further terms are of `polynomially' smaller order than unity.

With this, the relative error of $p_\mathrm{diff}$ with respect to the
random walk expressions can be calculated to yield
\begin{equation}
\Delta=\frac{\ln B-2(\ln 2-\gamma)}{\ln(a/W)+2(\ln 2-\gamma)}\,.
\end{equation}
Since $2(\ln 2-\gamma)\approx0.23$ and taking into account the values
of $B$ this means that in the large lattice regime, the diffusion
model encounter probability (and in its trail the sweeping rate and
the recombination efficiency) systematically exceeds that given by the
random walk result.  The discrepancy vanishes in the continuum limit
$S\to\infty$, however it only does so like $1/\ln(a/W)$ or as $1/\ln
S$ (since $SW/a=\const$).

\subsection{Small lattices}
The earlier results to compare the $p_\mathrm{rw}$ asymptotics against
are as follows: For small lattices, we obtained the diffusion model
expression
\begin{equation}
p_\mathrm{diff}\approx1-\frac{SW}{4 g a}\left[\ln (S/g)-1\right]\,,
\end{equation}
with $g$ as defined in the previous Section.  The heuristic random
walk derivation yielded
\begin{equation}
  p_\mathrm{rw,heur}\approx 1-\frac{CSW}{\pi a}\ln[BC/\pi\cdot S]\,.
\end{equation}
Including the factor $c_1$ occurring in $p_\mathrm{rw}$
of~\eqref{small-lattice-prw} in the relation established between $g$
and $C$, we observe that it satisfies 
\begin{equation}
  \label{constant}
  c_1=\frac{1}{4g}=\frac{C}{\pi}.
\end{equation}  
Thus once again, the pre-factors as well as the functional dependence
of all three expressions agree.

Turning to the pre-factor \emph{inside} the logarithm, we have to
resort to the numerical values of $\ln c$ as given in
Section~\ref{sec:rw-approx}.  Starting with the $p_\mathrm{rw,heur}$
expression, numerical evaluation of the corresponding $\ln BC/\pi$
yields numbers of $0.934711656$ and $1.196335728$ for type (a) and (b)
lattices, respectively.  While not in good agreement with the values
for $\ln c$, this is still reasonable: Apart from the nature of the
$N_\mathrm{dis}$ expansion, the ensuing \emph{heuristic} derivation
asserts that the walker does not die until it has traversed all $S$
sites, and includes inversion of the transcendental equation
$N_\mathrm{dis}(t)=S$ which is approximated by one iteration.  It is
the very pre-factor just mentioned that suffers from this
approximation.  In contrast, the analogous factor in the diffusion
model result reads $-\ln g-1$, and numerical evaluation provides
$-0.758435525\dots$ and $-0.902276561\dots$ for types (a) and (b),
respectively, both completely off.

For $S\to\infty$ with $SW/a=\const$ we will eventually still leave the
small lattice regime as $p$ drifts away from unity, cf.\ our earlier
remark regarding the used expansion.  To measure the error of the
diffusion model, it is hence more appropriate to compare the
complements $1-p$ of the two models, leading to a relative deviation
\begin{equation}
\Delta'=-\frac{\ln(cg\e )}{\ln(cS)}<0
\end{equation}
of the diffusion model outcome relative to that of the random walk.
Obviously, again the diffusion model delivers an encounter probability
systematically \emph{larger} than obtained from the random walk model,
and again, this discrepancy vanishes only logarithmically with the
system size $S\to\infty$.

\subsection{Results}\label{sec:comparison-results}

Besides the interest in the exact encounter probability of
the random walk $p_\mathrm{rw}$ and its limiting behavior for the two
regimes, we now have established that its asymptotics differ in
logarithmic terms from that of the continuum model $p_\mathrm{diff}$.
While the difference between discrete model results and those of the
continuum model vanishes in the true continuum limit $S\to\infty$,
$W/a\to\infty$ with $SW/a=\const$, it does so only logarithmically in
the system size $S$: This slow convergence is an inevitable direct
consequence of the marginality of spatial dimension two for the random
walk and diffusion, unfortunate for the comparison of discrete-space
simulations to analytic results.

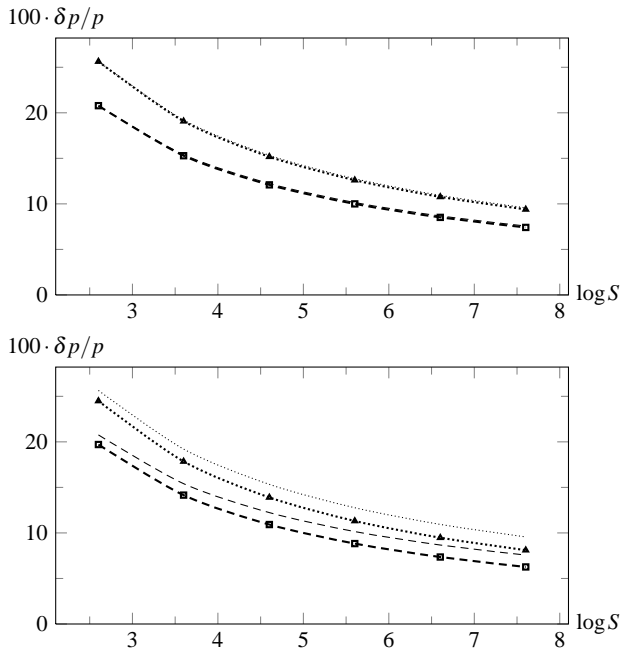
\begin{figure}
  \beginpgfgraphicnamed{graphic1a}%
  \begin{tikzpicture}
    \begin{axis}[y filter/.code={\pgfmathmultiply{#1}{-1}},
      xlabel=$\log S$,ylabel=$100\cdot\delta p/p$,
      width=\columnwidth,height=5cm,ymin=0,smooth,
      y tick label style={/pgf/number format/precision=0}]
      \addplot[densely dashed] plot file {dat/DeltaP_sq-3.dat};
      \addplot[densely dotted] plot file {dat/DeltaP_tr-3.dat};
      \addplot[densely dashed,thick,mark=square] plot file {dat/diffvsrwP_sq-3.dat};
      \addplot[densely dotted,thick,mark=triangle] plot file {dat/diffvsrwP_tr-3.dat};
    \end{axis}
  \end{tikzpicture}
  \endpgfgraphicnamed
  \\
  \beginpgfgraphicnamed{graphic1b}%
  \begin{tikzpicture}
    \begin{axis}[y filter/.code={\pgfmathmultiply{#1}{-1}},
      xlabel=$\log S$,ylabel=$100\cdot\delta p/p$,
      width=\columnwidth,height=5cm,ymin=0,smooth,
      y tick label style={/pgf/number format/precision=0}]
      \addplot[densely dashed] plot file {dat/DeltaP_sq-2.dat};
      \addplot[densely dotted] plot file {dat/DeltaP_tr-2.dat};
      \addplot[densely dashed,thick,mark=square] plot file {dat/diffvsrwP_sq-2.dat};
      \addplot[densely dotted,thick,mark=triangle] plot file {dat/diffvsrwP_tr-2.dat};
    \end{axis}
  \end{tikzpicture}
  \endpgfgraphicnamed
  \caption{Relative difference (in percent) of $1-p_\mathrm{diff}$
    w.r.t.\ $1-p_\mathrm{rw}$ for $SW/(4a)=10^{-3}$ (top) and
    $SW/(4a)=10^{-2}$ (bottom, still a \emph{small} lattice), type (a)
    (dashed\,/\,squares), type (b) (dotted\,/\,triangles); thinner lines the prediction
    $-\Delta'$.}
  \label{fig:anadelta-}
\end{figure}

\begin{figure}
  \beginpgfgraphicnamed{graphic1c}%
  \begin{tikzpicture}
    \begin{axis}[xlabel=$\log S$,ylabel=$100\cdot\delta p/p$,
      width=\columnwidth,height=5cm,ymin=0,smooth,
      y tick label style={/pgf/number format/precision=0}]
      \addplot[densely dashed] plot file {dat/Delta_sq0.dat};
      \addplot[densely dotted] plot file {dat/Delta_tr0.dat};
      \addplot[densely dashed,thick,mark=square] plot file {dat/diffvsrw_sq0.dat};
      \addplot[densely dotted,thick,mark=triangle] plot file {dat/diffvsrw_tr0.dat};
    \end{axis}
  \end{tikzpicture}
  \endpgfgraphicnamed
  \\
  \beginpgfgraphicnamed{graphic1d}%
  \begin{tikzpicture}
    \begin{axis}[xlabel=$\log S$,ylabel=$100\cdot\delta p/p$,
      width=\columnwidth,height=5cm,ymin=0,smooth,
      y tick label style={/pgf/number format/precision=0}]
      \addplot[densely dashed] plot file {dat/Delta_sq1.dat};
      \addplot[densely dotted] plot file {dat/Delta_tr1.dat};
      \addplot[densely dashed,thick,mark=square] plot file {dat/diffvsrw_sq1.dat};
      \addplot[densely dotted,thick,mark=triangle] plot file {dat/diffvsrw_tr1.dat};
    \end{axis}
  \end{tikzpicture}
  \endpgfgraphicnamed
  \caption{Relative difference (in percent) of $p_\mathrm{diff}$
    w.r.t.\ $p_\mathrm{rw}$ for $SW/(4a)=10^{0}$ (top, already a
    \emph{large} lattice) and $SW/(4a)=10^1$ (bottom), type (a)
    (dashed\,/\,squares), type (b) (dotted\,/\,triangles); thinner lines the prediction
    $\Delta$.}
  \label{fig:anadelta+}
\end{figure}
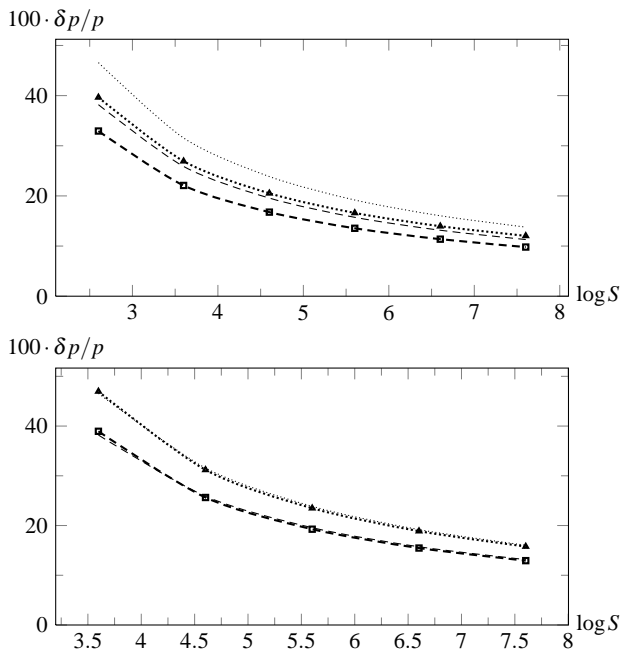

We finally plot (Figures~\ref{fig:anadelta-} and~\ref{fig:anadelta+})
the relative discrepancy of the \emph{exact} results of the diffusion
model $p_\mathrm{diff}$ with respect to that of the random walk
$p_\mathrm{rw}$ per cent, i.e.,
$100\cdot(p_\mathrm{diff}-p_\mathrm{rw})/p_\mathrm{rw}$ for the large
lattice regime, substituted by the relative difference of the
complements ($1-p$) for the small lattice plots.  Inside a single
plot, we sample lattice sizes determined by the lengths $L_1=L_2=L$
being the closest integers to give $S=4\cdot10^2,\,4\cdot10^3\dots$,
viz.\ $L=20$, $63$, $200$, $632$, $2000$, $6325$.  Simultaneously,
$\log(a/W)=2,\,3\dots$ increases with lattice size, such that the
`regime' parameter $SW/(4a)$ stays nearly constant and we properly
approach the continuum limit.  To compare, we also plot the error
estimates $\Delta$ and $\Delta'$ as appropriate for the given regime.

In both regimes the error predictions nicely agree with the
discrepancy of the exact results, with increasing precision the
further we go into a certain regime.  This also verifies that the
asymptotic results used to derive the estimates are correct, and that
the discrepancy is not due to any relevant terms erroneously omitted.
The error generally shows the predicted slow logarithmic decrease in
the system size.  Moreover, one can clearly see that it is
considerable for all lattices, even for the largest ones, and the
error even increases the further we enter one regime.

\section{Simulations}\label{sec:simulations}

We already mentioned that in many applications, the full microscopic
kinetics is simulated within the Monte Carlo approach.  Thence here we
quantitatively compare our analytic predictions for the encounter
probability to the corresponding simulation results.  It is fairly
easy to simulate both, the stochastic two-particle system for which we
defined the encounter probability $p$, but also the `full'
diffusion-reaction system including continuous stochastic deposition
of new random walkers, yielding results for the recombination
efficiency of this process.

We use a standard kinetic Monte Carlo algorithm to keep track of
individual atoms deposited onto, hopping around on and desorbing from
the lattice.  As in the random walk description, the lattice is
homogeneous, with periodic boundaries.  All waiting times for events
are stochastically distributed according to an exponential
distribution with the average waiting time given by the inverse rate
of the corresponding process.  In the simplified case of the
two-particle simulation we consider two independently moving and
desorbing atoms in continuous time -- though with an arbitrarily
rescaled time step permitted by the homogeneous environment.  This is
the simplest implementation of the continuous-time random walk (CTRW)
introduced by \cite{Montroll65}.

The algorithm respects the \emph{Langmuir-Hinshelwood rejection}
described earlier, i.e., atoms impinging on top of occupied sites are
repelled.  Hence for faithful comparison between simulation and random
walk theory, in this whole Section we will refer to $\tilde
p_\mathrm{rw}$ instead of $p_\mathrm{rw}$.

Let us first relate what follows to earlier work.  The simulation of a
single moving (and possibly desorbing) atom that is to meet a fixed
immortal target site in a homogeneous environment (hence the situation
used in deriving $p_\mathrm{rw}$) obviously obliterates any notion of
continuous time and can be fully described in terms of \emph{steps} of
the walk.  It is therefore a truism that any correctly working
microscopic Monte Carlo simulation of such a system has to converge to
$p_\mathrm{rw}$ if we sample enough individual trials.  For the case
without desorption (where the interest is not in the probability of an
encounter, but rather in the statistics of the number of steps this
takes), the predictions of \cite{Montroll69} have been confirmed by
Monte Carlo simulations in \cite{Hatlee80}, where the effect of
alternative boundary conditions of the finite lattice is also
examined.

Now if both reactants can move and desorb, it makes sense to simulate
them in continuous time to let all events happen in an ordered
fashion, and this is what we have done in all simulations we refer to
herein.  The equivalence of both scenarios (two walkers vs.\ one
walker, and regarding the encounter probability) was clear in the
stationary diffusion model, due to its continuum nature, the absence
of any time, and its dealing with a ``concentration'' that is obtained
by averaging over random walk realizations.  For the single instance
of the CTRW we note that (in our homogeneous setting) we can still
treat one of the two walkers as fixed: If it is to make a move, we may
simply re-label lattice sites accordingly and let the other walker
perform an appropriate step instead.  If it dies, the trial is ended
as well as if the other walker had died.  Hence the system can equally
be described as a single walker that is to meet a target site, hopping
and dying at twice the rates of each of the two original random walks,
and still with exponentially distributed waiting times.  This change
of rates does not alter the survival probability per step and thence
does not change the encounter probability.  Moreover, while the number
of steps performed in a given time is a stochastic quantity for the
CTRW (as opposed to the discrete-time random walk, their relation for
a huge class of waiting time distributions examined in
e.\,g.~\cite{Bedeaux71}), the time passed never matters for the
encounter probability, so that a description in terms of steps only is
fully justified.

Indeed, we can report excellent agreement of the encounter probability
$\tilde p_\mathrm{rw}$ obtained from the random walk model with that
found in Monte Carlo simulations, $\tilde p_\mathrm{mc}$, see
Appendix~\ref{app:sims-pmc-vs-prw} for details.  This holds for
arbitrary parameters $S$ and $W/a$, as well as for both types of
lattice, and gives credence to the correctness of our simulation
method.

To obtain $\tilde p_\mathrm{mc}$ we sampled $N=10^6$ individual trials
(except for the largest $S$ and $\ell$,
where we chose $N=10^5$ due to time constraints) of the Monte Carlo
simulation of two atoms moving in continuous time.  Such a repeated
sampling of independent \textsc{Bernoulli} trials with (unknown) exact
success probability $p$ (corresponding to recombination in our
context) yields an absolute standard deviation of the outcome of
$\sigma=\sqrt{p(1-p)/N}$.  Thus $\tilde\sigma=\sigma/p$, evaluated
using the (known) values for $\tilde p_\mathrm{rw}$, is the expected
\emph{relative} standard deviation for simulations.  Obviously
$\tilde\sigma$ becomes fairly large for small $p$, i.e., approaching
the large lattice regime.

The binomial distribution for the number of recombinations with fixed
success probability $p$ of the individual trial and for the number of
samples $N\to\infty$ tends to a Gaussian distribution, so the
distribution for the numerically found $\tilde p_\mathrm{mc}$ should
tend to a normal distribution with the above standard deviation.
Roughly a third of all results are then expected to lie outside a
corridor of half width $\tilde\sigma$ due to fluctuations; counted
over all simulation data points for the quadratic lattices 
we find precisely $26/78$ values
outside this range.  The largest deviations from $\tilde
p_\mathrm{rw}$ are of the order of less than $3\tilde\sigma$, with the
largest absolute deviations smaller than $7\%$.  Furthermore there is
no systematic over- or underestimate of the theory.  Similar results were
obtained for rectangular lattices (see Appendix~\ref{app:sims-pmc-vs-prw}).  
We conclude that the simple random walk result describes the continuous-time
simulations to excellent accuracy.  On a side note, this shows that
the diffusion model results would completely fail to describe
simulation results, most pronounced in the large lattice regime.

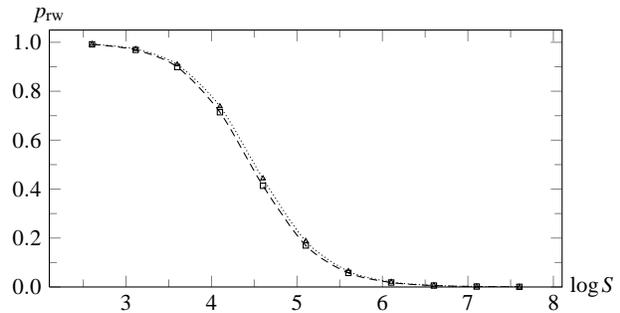
\begin{figure}
  \beginpgfgraphicnamed{graphic1e}
  \begin{tikzpicture}
    \begin{axis}[width=\columnwidth,height=5cm,
      xlabel=$\log S$,ylabel=$p_\mathrm{rw}$,
      ymin=0,ymax=1.05,smooth,try min ticks=5,
      y tick label style={/pgf/number format/precision=1}]
      \addplot[densely dashed,mark=square]
      table[x index=0,y index=1] {dat/prwlogS.dat};
      \addplot[densely dotted,mark=triangle]
      table[x index=0,y index=2] {dat/prwlogS.dat};
    \end{axis}
  \end{tikzpicture}
  \endpgfgraphicnamed
  \caption{Random walk encounter probability $p_\mathrm{rw}$ as a
    function of the (logarithmic) grain size for $W/a=10^{-5}$, and
    for a type (a) (dashed\,/\,squares) or a type (b)
    (dotted\,/\,triangles) lattice,
    respectively.}\label{fig:prw-lattice-types}
\end{figure}
The excellent agreement between simulations and random walk theory
allows us to revisit the effect of the lattice type on the
recombination efficiency, which was discussed in~\cite{Lohmar06} with
reference to the Monte Carlo simulations reported in~\cite{Chang05}.
We have seen above in Section~\ref{sec:comparisons} that, to leading
order, the lattice type enters the encounter probability through the
multiplicative constant~\eqref{constant}, which differs by about
$15\%$ between the square and triangular lattices.  The effect
obtained using the full, exact expression for $p_\mathrm{rw}$ is of
the same order or smaller, as illustrated in
Figure~\ref{fig:prw-lattice-types}.  Since neither the sweeping rate $A$
nor the recombination efficiency depend more strongly on the lattice
type than $p$ itself~\cite{Lohmar06}, we may conclude that significant
lattice type effects cannot be expected in simulations that faithfully
represent the reaction-diffusion model treated in this paper.  The
discrepancy with the results of~\cite{Chang05}, where efficiencies on
square and triangular lattices were found to differ by a factor of 2,
thus remains.

\section{Shape of the lattice}
\label{sec:shape}

As was mentioned in \cite{Lohmar06,Krug03}, the diffusion model
results for the encounter probability can be compared to those
obtained in a completely analogous way for a flat disc with a
reflecting outer boundary and a fixed absorbing target in its center.
A detailed analysis (again also accounting for encounters ``by
deposition'') shows that upon natural identification of the
parameters, in both regimes the functional form and pre-factors
coincide, but while in the ``large-disc'' regime the asymptotics
reproduces the sphere result~\eqref{pdiff-large-sphere} exactly
\emph{including the logarithmic factor} (or coequal $\mathcal{O}(1)$
terms), it slightly differs in the ``small-disc'' regime, where
\begin{equation}
p_\mathrm{diff,disc}\approx1-\frac{SW}{4 g a}\left[\ln (S/g)-3/2\right]\,,
\end{equation}
with the lattice factor $g$ defined before.  Numerical evaluation
shows that there is only a small (of the order of a few percent)
difference of both models in between these limiting cases.

This re-assures our conviction that curvature effects should not
matter with $S\gg1$: The basic reason is that the radius of curvature
(in units of lattice spacing) is of the order of the system size.  The
absence of any effect is obvious for the large-lattice regime, when a
random walk typically does not travel a distance long enough to feel
the radius of curvature, $\ell\ll\sqrt{S}$.  But as soon as
$\ell\sim\sqrt{S}$, nearly the whole lattice is swept by the walk
anyway, the small fraction not explored not depending on curvature,
but on the failure of \emph{locally} dense exploration.

Note that the difference of the logarithmic factors of sphere and disc
in the small-lattice regime appears in the same order of magnitude as
the discrepancy between lattice and continuum models in
Section~\ref{sec:comparisons}, although the former models (in contrast
to the latter ones) substantially differ in topology and boundary
conditions.  While at some point we had considered curvature and
connectivity effects responsible for the discrepancy between the
diffusion model result and that obtained for the random walk, this
comparison strongly (and quantitatively) suggested otherwise.

It is clear, however, that boundary conditions and the general shape
of the lattice should have \emph{some} effect on the random walk
properties (at least in certain parameter regimes), and possibly on
the encounter probability $p$.  Related questions concerning the
importance of confined geometries, without desorption and
correspondingly focusing on mean first-passage times instead of
first-passage probabilities, have been examined in detail in a recent
remarkable series of papers \cite{Condamin05,Condamin06,Condamin07}.
Moreover, as we have seen, the peculiarities of spatial dimension two
are very important for the features of $p$ -- what happens then, is
one naturally ensuing question, if we distort the lattice shape in such
a way that it becomes \emph{effectively} one-dimensional?

In view of the line of thought of this article we focus here on the
analysis of a torus lattice (periodic and rectangular) with distorted
aspect ratio, that is, with different lengths $L_1$ and $L_2$ in the
two directions.  We will refer to the case $L_1\neq L_2$ as
\emph{rectangular}, while we call $L_1=L_2$ a \emph{quadratic}
lattice, the latter not to be mixed up with the term \emph{square}
that we use exclusively to label a certain lattice type (i.e., its
internal structure) as opposed to its shape or geometry.

Let us therefore refine our notion of small and large lattice regimes.
We still assume all three lengths $L_1$, $L_2$, $\ell \gg1$
throughout.  Without any loss of generality, let $L_2\leq L_1$, but we
might further specialize to $L_2\ll L_1$.  Ordering lengths, we are
then left with three (instead of our former two) asymptotic regimes
defined by the ordering of length scales:
\begin{itemize}
\item[$\bullet$] $1\ll \ell \ll L_2\leq L_1$.
\item[$\bullet$]  $1\ll L_2\ll \ell \ll L_1$.
\item [$\bullet$] $1\ll L_2\leq L_1\ll \ell $.
\end{itemize}
The first case corresponds to the earlier large-lattice regime; the
third case to small lattices.  The intermediate regime in which the
walk easily sweeps one dimension but is typically short compared to
the other is a new property.  We define the \emph{aspect ratio} as
$\mu:=L_2/L_1\in(0,1]$.

We will dismiss the large-lattice case for most of the sequel.  In
this regime, boundaries are effectively not felt by the random walker,
and the appropriate double integral limit of~\eqref{general-prw}
yields a result that only depends on the \emph{total} number of sites
$S$ (and no longer on $L_{1,2}$), as given in
Section~\ref{sec:prw-large-lattice}.

\subsection{Overall behavior}
We keep $S$ and $\xi$ constant and show the encounter probability
$p_\mathrm{rw}$ on a type (a) square lattice as a function of (the
logarithm of) the aspect ratio $\mu$ for both, one absolutely small
($S=4\times10^2$) and one absolutely large lattice ($S=4\times10^6$).
In Figure~\ref{fig:plot-ar-s}, we start in the small-lattice regime
for the quadratic case (rightmost in the plot), whereas
Figure~\ref{fig:plot-ar-l} starts in the large-lattice regime.  There
is no qualitative difference between the two lattice types with
respect to any plot.
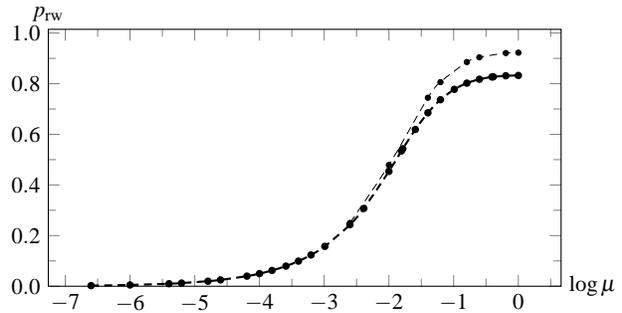
\begin{figure}
  \beginpgfgraphicnamed{graphic2a}
  \begin{tikzpicture}
    \begin{axis}[title style={at={(0.5,1.1)},anchor=center},
      xlabel=$\log \mu$,ylabel=$p_\mathrm{rw}$,
      width=\columnwidth,height=5cm,ymin=0,smooth,
      y tick label style={/pgf/number format/precision=1},
      legend style={at={(0.03,0.93)},anchor=north west}]
      \addplot[mark=*,very thin,densely dashed] plot file {dat/prw2s-mu.dat};
      \addplot[mark=*,thick,densely dashed] plot file {dat/prw6s-mu.dat};
    \end{axis}
  \end{tikzpicture}
  \endpgfgraphicnamed
  \caption{Encounter probability (on a square lattice) as a function
    of $\log \mu$, for fixed sizes $S=4\times10^6$ (thick) and
    $S=4\times10^2$ (thin), respectively, and $SW/(4a)=10^{-2}$ in
    both cases.}\label{fig:plot-ar-s}
\end{figure}
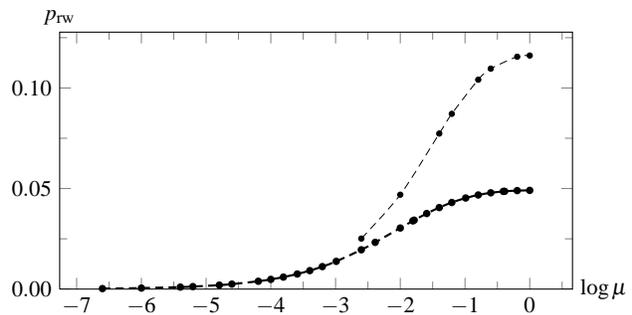
\begin{figure}
  \beginpgfgraphicnamed{graphic2b}
  \begin{tikzpicture}
    \begin{axis}[title style={at={(0.5,1.1)},anchor=center},
      xlabel=$\log \mu$,ylabel=$p_\mathrm{rw}$,
      width=\columnwidth,height=5cm,ymin=0,smooth]
      \addplot[mark=*,thick,densely dashed] plot file {dat/prw6l-mu.dat};
      \addplot[mark=*,very thin,densely dashed] plot file {dat/prw2l-mu.dat};
    \end{axis}
  \end{tikzpicture}
  \endpgfgraphicnamed
  \caption{As Figure~\ref{fig:plot-ar-s} with
    $SW/(4a)=1$.}\label{fig:plot-ar-l}
\end{figure}

In all cases the encounter probability shows a strong monotonic
decline upon distortion (i.e., moving left in the Figures), though
with a less pronounced shape and ending at a fair fraction of its peak
value for the absolutely small lattice.  We will first analyze several
aspects of this transition, finally putting our findings together in a
qualitative explanation.

\subsection{Extreme distortion}\label{sec:extreme-dist}

We start with a simple observation.  First, let us use the
two-dimensional result for $p_\mathrm{rw}$ on a type (a) square
lattice, and let one lattice dimension shrink to one lattice unit only
($S=L_1L_2=L\cdot1$).  What we get from~\eqref{general-prw} is
\begin{equation}
  p_\mathrm{rw}^{-1}=
  \sum_{m=0}^{S-1}\frac{1-\xi}{1-(\xi/2)
    \left(\cos(2\pi m/S)+1\right)}\,,
\end{equation}
where we have not yet used any summation results.  Second, we
immediately start with a genuine one-dimensional lattice, and then the
sum for an unbiased walk (different structure function!) reads
\begin{equation}
  p_\mathrm{rw,\,1d}^{-1}=
  \sum_{m=0}^{S-1}\frac{1-\xi}{1-\xi
    \cos(2\pi m/S)}\,,
\end{equation}
which is obviously different from the former limit.  We see that
$p_\mathrm{rw,\,1d}>p_\mathrm{rw}$ for equal number of adsorption
sites $S$ and equal $\xi$.

The explanation of this result of an ``extremely distorted'' situation
provides some insight for the general problem.  In the two-dimensional
version, the walker still performs steps in the direction in which
there is only one lattice unit length (of the \emph{periodic}
lattice), and one can imagine it as walking around on a very `thin'
torus with one dimension completely ``wrapped up''.  Contrary to this,
the truly one-dimensional walk only performs steps in the direction in
which the lattice is extended.  The net effect is that the
two-dimensional walker wastes on average half the number of steps it
takes by walking in the `wrong' dimension and not coming any closer to
the target site.  In fact, this suffices to re-gain one result from
the other heuristically: The 2d-walk corresponds to a 1d-walk with the
same desorption rate $W$, but with the undirected hopping rate
effectively halved by the useless waiting steps, $a\to a/2$.  This
change translates to $(1-\xi)/\xi\to2(1-\xi)/\xi$, and minimal
manipulation of the two expressions given above shows that this casts
the 1d-walk result into that for the two-dimensional walk
(see~\cite{Montroll73} for a related discussion).

\subsection{Small lattices and \boldmath{$L_1\neq L_2$}}
\label{sec:montroll-p-star}
For the following analysis, we generalized the asymptotic result for
$P^*(\vec 0;\xi)$ of \cite{Montroll69,denHollander82} to the case
$L_1\neq L_2$, at least in the vicinity of the square lattice.  To
this end, one mainly has to diligently separate the different $L_j$s
in the derivation and to check that the individual steps continue to
hold.  Based on the generalized \emph{exact} expressions given in
Appendix~\ref{app:one-sum}, we obtain
\begin{equation}\label{P-star-ar}
\begin{split}
&\qquad P^*(\vec 0;\xi) 
= \frac1{L_1L_2(1-\xi)} + \frac{\ln L_1}{r\pi(1-2q_0)} \\
&+ \frac{\frac{L_2}{3L_1}+
  \frac1{r\pi}[2\gamma+2\ln(2/\pi)-\ln(1+\eta)]+S_3^{(0)}/r}{2(1-2q_0)} \\
&+
\frac{-\frac13+\frac{(3\eta-1)\pi}{36r}\frac{L_2}{L_1}+S_3^{(1)}/r}{2(1-2q_0)L_1L_2} \\
&+ \order(L^{-4})+\order(1-\xi)^{1/2},
\end{split}
\end{equation}
where, first of all, terms are arranged in orders of $\sqrt{1-\xi}$,
as this is the smallest quantity, and, inside, according to orders of
$L_{1,2}$. The parameter $q_0$ is the probability a step is directed
into a particular lattice unit direction, $q_1$ and $q_2$ denote
transition probabilities into the diagonal directions, cf.\
Appendix~\ref{app:one-sum}.  Furthermore, we adopted the definitions
\begin{equation}\begin{aligned}
r&=\frac{2[(q_0+2q_1)(q_0+2q_2)]^{1/2}}{1-2q_0}\,\\
\eta&=\frac{q_0(1-2q_0)}{(q_0+2q_1)(q_0+2q_2)}-1\,,
\end{aligned}
\end{equation}
which take the values $r=1=\eta$ for the square lattice case (a), and
$r=\sqrt{3}/2$, $\eta=1/3$ for the triangular lattice case (b).  The
only parts in the above expression that have to be explicitly
evaluated for the two lattice types are the sum contributions
$S_3^{(0,1)}$.  Unfortunately, their generalization to $L_1\neq L_2$
is reasonably feasible only for the square lattice case (a), on which
we now focus and where
\begin{equation}\label{P-star-sums}\begin{aligned}
S_3^{(0)} &= \frac{4}{\pi}
\Bigl(\e^{-2\pi L_2/L_1}+\frac32\e^{-4\pi L_2/L_1}
+\frac43\e^{-6\pi L_2/L_1}+\dots\Bigr),\\
S_3^{(1)} &= -\frac{4\pi}{3}
\left(\frac{L_2}{L_1}\right)
\Bigl(\e^{-2\pi L_2/L_1} +3\e^{-4\pi L_2/L_1}\\
&\qquad+4\e^{-6\pi L_2/L_1}+\dots\Bigr)\\
&\quad +\frac{8\pi^2}{3}
\left(\frac{L_2}{L_1}\right)^2
\Bigl(\e^{-2\pi L_2/L_1}+6\e^{-4\pi L_2/L_1}\\
&\qquad+12\e^{-6\pi L_2/L_1}+\dots\Bigr).
\end{aligned}
\end{equation}

Due to the nature of the involved approximations, the expansion is
asymmetric in the lattice lengths.  In the form given above, $L_2$ is
the limit of the `inner' sum that is explicitly and exactly evaluated
(see Appendix~\ref{app:one-sum}), while the other length $L_1$ appears
as the upper limit in several Euler-MacLaurin formulae in the
derivation.  Therefore we expected the result would fit best for the
case where $L_2\leq L_1$.  This is the ordering we will consistently
assume throughout the following.

The detailed assumptions on the length scales that enter the result
are as follows: First, we need $L_1L_2(1-\xi)\ll1$ for the general
form of the $P^*$ expansion to be meaningful, cf.~\cite{Montroll69},
Eq.\ B35.  Further, the result relies on the fact that the dimension
for which the sum is explicitly evaluated is small, $L_2\ll \ell
=\sqrt{\xi/(1-\xi)}\approx1/\sqrt{1-\xi}$, necessary for the used
expansion of the first summand to converge (\emph{ibid.} Eqs.\ B11,
B12), while the other is absolutely large $L_1\gg1$ (for applicability
of the \textsc{Euler-MacLaurin} formula).  Lastly, $L_2/L_1$ must not
be \emph{too} small (owed to convergence in Eqs.\ B28 ff.\
\emph{ibid.}, also see \eqref{P-star-sums}).  This set of conditions
applies to the small-lattice regime whenever $L_2/L_1$ does not become
too small; but it is also satisfied for the intermediate regime
provided that \emph{additionally}, $L_1L_2(1-\xi)\ll1$.

It is fairly obvious that under certain conditions we may exchange the
role of the lengths $L_2<L_1$ in the derivation, such that the
summation is carried out for the larger length $L_1$ instead.  Thence
the results \eqref{P-star-ar} and \eqref{P-star-sums} will remain
valid upon swapping all occurrences of $L_1$ and $L_2$ if the
aforementioned conditions remain satisfied; we refer to this as the
`swapped-lengths' version.  As then, $L_1\ll\ell $ is necessary, we
are restricted to the small-lattice regime; the remaining condition is
equivalent to $L_1/L_2^3\ll1$, as follows from careful inspection of
the B28 derivation of \cite{Montroll69}.

A consistency check between the result as it reads above and the
swapped-lengths version has to take into account that it can only hold
in the small-lattice regime, and that the orders of several terms
change.  Numerical comparison of both versions and comparisons with
the evaluation of the exact result $p_\mathrm{rw}$ show that, indeed,
the swapped-lengths version reproduces the exact result more
accurately than the form given above, as it circumvents the
aforementioned convergence issue.  This is only a minute problem
however, and both versions agree and are in good agreement with the
exact result (for the square lattice) over a certain range of
distortions (checked for $\mu=L_2/L_1=1/4\dots1$ for the small-lattice
regime with $S=4\cdot10^6$, $a/W=10^8$, the agreement of the
swapped-lengths version with the exact result extending farther down
to $\mu\sim1/10$).

\subsection{Small distortion}

We now want to approach the question of the aspect ratio influence not
from the extreme case, but in contrast starting to deform a quadratic
lattice.  Concretely, we ask when (and to what effect) the random walk
feels the global change if we slightly distort a quadratic lattice of
$S=L^2$ sites, keeping $W/a$ and $S$ fixed.  The latter condition is
necessary since (judging from previous results) the effect of varying
$S$ is dominant otherwise; this hinders us from simply distorting the
lattice lengths to ``nearby'' integer dimensions ($(L-1)(L+1)=S-1$).

Therefore we proceed as follows: We treat $p_\mathrm{rw}$ as a
function $p_\mathrm{rw}(\mu,S,\xi)$, keep the latter two arguments
fixed, and assume that the aspect ratio $\mu$ can be varied
continuously.  Any statements about the local behavior of that
function around the quadratic shape value $\mu^*$ remain sensible as
long as higher-order terms do not restrict their validity to intervals
too small.  We do not believe this to be the case, basically because
we cannot conceive of any mechanism that could render the behavior of
$\left.p_\mathrm{rw}\right\rvert_{S,\,\xi=\mathrm{const},\,
  \mu\leq\mu^*}$ non-monotonic.

The effect of a small distortion could depend on the regime we are in.
We now focus on the small-lattice regime, since we expect aspect ratio
effects to be most prominent there, and since for this case, we could
obtain the general asymptotic behavior in
Section~\ref{sec:montroll-p-star}.  Note that starting from a square
lattice with the question at hand rules out the intermediate regime,
and that of large lattices was argued not to be of much interest
above.

Now $\d^2p/\d\mu^2
=[S(1-\xi)]^{-1}\frac{\d^2(P^*(\vec0;\xi))^{-1}}{\d\mu^2}$, and at the
extremal point, $\frac{\d^2(P^*(\vec0;\xi))^{-1}}{\d\mu^2} =
-(P^*(\vec0;\xi))^{-2}\frac{\d^2P^*(\vec0;\xi)}{\d\mu^2}$, since the
first derivative vanishes ($\left.\d p/\d\mu\right\rvert_{\mu=\mu^*}
=0$ due to $L_1\leftrightarrow L_2$ symmetry of $p$).  For our
purposes we only need to find the sign of the second $P^*(\vec0;\xi)$
derivative.  To this end, we use the approximation of
Section~\ref{sec:montroll-p-star}, supposing that the dominant
contributions to $P^*(\vec0;\xi)$ are also dominating the sign of the
second derivative.  As a function of $\mu$, \eqref{P-star-ar} reads
\begin{equation}
  \begin{split}
    P^*(\vec0;\xi) &= \mathrm{const} + \frac{\ln\sqrt{S/\mu}}{r\pi(1-2q_0)}
    + \frac{\mu/3+S_3^{(0)}/r}{2(1-2q_0)}\\
    &+ \frac{\frac{(3\eta-1)\pi}{36r}\mu+S_3^{(1)}/r}{2(1-2q_0)S}
    + \order\left(L^{-4},(1-\xi)^{1/2}\right),
  \end{split}
\end{equation}
and therefore in the second derivative, only the $S_3$ contributions
and those from the square root term remain, viz.\
\begin{equation}
  \frac{\d^2P^*(\vec0;\xi)}{\d\mu^2}
  \approx \frac{\frac1{\pi}\mu^{-2}
  + \d^2(S_3^{(0)}+S_3^{(1)}/S)/\d\mu^2}{2r(1-2q_0)}\,.
\end{equation}
The $S_3^{(1)}$ contribution is irrelevant due to the $S^{-1}$ factor, and
\begin{equation}
  \begin{split}
    \frac{\d^2S_3^{(0)}}{\d\mu^2}
    &= 16\pi
    \Bigl( \e^{-2\pi\mu} + 6 \e^{-4\pi\mu}
    + 12 \e^{-6\pi\mu} +\dots \Bigr)
    \,,
  \end{split}
\end{equation}
where (as before), we only treat case (a) with $r=1=\eta$.  Numerical
evaluation of this term at $\mu=1$ yields approximately $0.095$.
Thence, $\frac{\d^2P^*(\vec0;\xi)}{\d\mu^2}>0$, and consequently,
$\d^2p/\d\mu^2<0$: the encounter probability of the random walk in the
small-lattice regime has a maximum for the quadratic lattice, and it
decreases with any distortion from this shape.\footnote{We checked by
  numerical evaluation that $P^*(\vec0;\xi)$ is monotonic, or
  equivalently, that the first derivative of $P^*(\vec0;\xi)$ indeed
  does not change its sign from $\mu=1$ to $\mu=1/4$, corresponding to
  the distortion $L_1\to2L_1,\, L_2\to\frac12L_2$ from a quadratic
  lattice with even lengths $L=L_1=L_2$.  Here we used the
  swapped-lengths version of~\eqref{P-star-ar} to avoid convergence
  issues and improve compliance with the exact result as detailed in
  Section~\ref{sec:montroll-p-star}; the necessary conditions for this
  to be allowed are easily satisfied.}  This is the quantitative
foundation of our heuristic arguments in
Section~\ref{sec:ar-qualitative}.

\subsection{The quasi-one-dimensional limit}\label{sec:quasi-1d-limit}

Finally we want to clarify the nature of the transition to effective
one-dimensionality: What happens if the aspect ratio goes to infinity,
so that the macroscopic structure of the lattice becomes
quasi-one-dimensional?

First, we comment on the relationship between the regimes of the
one-dimensional and the distorted two-dimensional lattices.  In the
2d-large-lattice regime, $\ell $ is the smallest length scale; this
cannot be sensibly related to any one-dimensional situation.  It is
then clear that the large-lattice regime on a one-dimensional lattice
of length $L\gg\ell_\mathrm{1d}$ corresponds to the intermediate
regime of the two-dimensional situation, i.e., $1\ll L_2\ll\ell \ll
L_1$, the common feature being that exactly one lattice dimension is
much larger than the typical random walk.  The remaining small-lattice
regime of the two-dimensional case ($1\ll L_2\ll L_1\ll \ell $) has
its analogon in the small-lattice regime $1\ll L\ll\ell $ of the
one-dimensional case, in both cases defined by a diffusion length that
exceeds \emph{any} lattice dimension.

We now consider a scaling limit of the three lengths in the
(two-dimensional) small-lattice regime so that we can apply the result
of Section~\ref{sec:montroll-p-star}.  Since $\ell $ is the largest
length, we use $\ell\to\infty$ as the basic scale in powers of which
we express the $L_i$ behavior.  In the original form
\eqref{P-star-ar}, $L_2/L_1\to0$ is not permitted, whence we employ
the swapped-lengths version again.  Now let $L_1\sim\ell$ and
$L_2\sim\ell^{1/2}$ (for the latter, any exponent in $[1/3,1)$ can be
chosen, the lower bound presumably an artefact of the details of the
involved approximations only).  This implies
$L_1/L_2^3\sim\ell^{-1/2}\to0$, thus satisfying a central assumption
for the approximations, while $L_1/L_2\sim\ell^{1/2}\to\infty$, and
consequently, the $S_3^{(0),(1)}$ terms vanish as well.  Moreover,
$L_1L_2(1-\xi)\sim\ell^{-1/2}\to0$ and $L_1/L_2\cdot
L_1L_2(1-\xi)\to\mathrm{const}$.  This scaling describes a situation
where all lengths diverge, but the larger lattice dimension scales
with the random walk length, while the smaller length increases more
slowly to distort the lattice to vanishing aspect ratio $L_2/L_1$.

Putting all of this together we obtain
\begin{equation}\label{small-lattice-scaling}
p_\mathrm{rw}^{-1}\approx 1+L_1L_2(1-\xi)\left\{
\frac{\ln L_2}{r\pi(1-2q_0)}
+\frac{L_1/L_2}{6(1-2q_0)}\right\},
\end{equation}
where higher-order terms in the brackets were omitted.  Here, the
logarithmic correction terms (and coequal $\mathcal{O}(1)$ terms in
$P^*(\vec0;\xi)$) can be seen to die out, while the second correction
term actually approaches a constant.  Neglecting all higher-order
terms we thence have
\begin{equation}
p_\mathrm{rw}\approx 1-\frac{L_1^2(1-\xi)}{6(1-2q_0)},
\end{equation}
or for the natural choice of a square lattice ($q_0=1/4$), and with
$1-\xi \approx\ell ^{-2}$ to leading order,
\begin{equation}
p_\mathrm{rw}\approx 1-\frac{(L_1/\ell )^2}{3}.
\end{equation}
We compare this to the corresponding one-dimensional small-lattice
regime result~\eqref{p1d-small-lattice}, which implies (cf.\
Appendix~\ref{app:asy-one-dim})
\begin{equation}
p_\mathrm{rw,1d}\approx 1-\frac{(L/\ell_\mathrm{1d})^2}{6}.
\end{equation}
The proper scaling limit results in a crossover from logarithmic
correction terms in the full two-dimensional small-lattice regime to
the squared length ratio correction we found for the corresponding
genuinely one-dimensional small-lattice regime.  Only the larger
length (scaling as the random walk length) remains, while the smaller
one has disappeared from the result.  Note that the appropriate
rescaling of the random walk length,
$\ell_\mathrm{1d}\to\ell/\sqrt{2}$ (accounting for the splitting up of
the number of steps between the two dimensions, as explained in
Section~\ref{sec:extreme-dist} and in Appendix~\ref{app:asy-one-dim})
reproduces the exact numerical pre-factor of the leading correction
term, as it should.

\subsection{Discussion}
\label{sec:ar-qualitative}

We now explain in a coherent fashion the main effects which govern the
behavior of the encounter probability on a distorted lattice, as shown
in Figures~\ref{fig:plot-ar-s} and~\ref{fig:plot-ar-l}.  We continue
to keep $S$ and $\xi$ (or $\ell$) constant.

The total decline from the peak value of $p_\mathrm{rw}$ to its
minimum for the fully distorted $S\cdot1$ lattice strongly depends on
the absolute lattice size: While for the absolutely large lattice, the
probability dropped to less than $0.5\%$ of that in the quadratic case
peak, we still have $1/6$ of the peak value left when $L_2=1$ with the
absolutely small lattice.  The quadratic lattice has either $p\sim1$
or $p\sim (a/W)/S$, but upon maximal distortion it invariably ends
effectively one-dimensional (with halved hopping rate) in its
1d-large-lattice regime, such that $p_\mathrm{1d}\sim\sqrt{a/W}/S$,
see Appendix~\ref{app:asy-one-dim}.  Fixing the quadratic lattice
regime via $SW/a$ then implies that $p$ also stays fixed as a function
of $S$, and hence yields the ratio $p/p_\mathrm{1d}\propto S\sqrt{W/a}
=\sqrt{SW/a}\sqrt{S}\propto\sqrt{S}$ \emph{in both regimes}.  With
increasing length the two-dimensional walk misses many
sites~\cite{vanWijland97}, but still sweeps an area of order $\ell^2$
in the large-lattice limit, while the one-dimensionality of the
extremely distorted case (also in its large-lattice regime) changes
the $\ell$ power to the unfavorable.  Obviously, a transition from a
quadratic 2d-small-lattice into the 1d-small-lattice regime would only
yield minute corrections to near-perfect encounter probability.

As for the shape of the $p_\mathrm{rw}$ dropoff, there are two effects
which cooperate in \emph{decreasing} the encounter probability due to
distortion, related to the \emph{dynamics} and the \emph{initial
  conditions} of our problem.

Regarding the dynamics, i.e., the exploration of the surface, consider
the effect of `wasted steps' (described in
Section~\ref{sec:extreme-dist}) in the smaller lattice dimension.
Close to the quadratic shape there are no wasted steps, while close to
maximal distortion the effect is most pronounced, as steps in the
smaller direction are nearly useless indeed -- but even there the
effect is that of halving the hopping rate, or modifying the random
walk length $\ell$ by a factor of $\mathcal O(1)$.  Expecting only
power-law behavior of $p$ in the lengths, this hardly shows on our
logarithmic scale.  Motion in the smaller dimension $L_2$ \emph{does}
become `wasteful' once $\ell\ll L_2$, when more steps or further
reduction of $L_2$ do not help the walker to get closer to the target:
Its residence probability has already spread out in this direction,
and new territory can only be explored by stepping out in the other
dimension ($L_1$).  But in a regime where this substantially
\emph{worsens} the chance to reach the target, it is owed to extending
the larger lattice length only.  Hence the `dynamic' effect lies in
the changing ratios $L_{1,2}/\ell$ upon distortion, governing how fast
or to what extent the lattice is explored in each direction, and
eventually switching the (refined) regimes.

The second effect is `static' in that it depends on the initial
conditions and not on the dynamics of the system.  We always assume
deposition of the walker homogeneously distributed over the lattice.
Even when, in the small-lattice regime, the largest parts of the
lattice will be swept by most walks, it is still an obstacle for the
\emph{individual} walk if it starts further away from the target.  Now
as soon as the lattice becomes distorted, the complete distribution of
the walker's initial distance (and in particular its average) to the
target is shifted to larger distances, as one can convince oneself of
with a simple sketch.  For the large-lattice regime, a region of an
extension $\ell$ surrounding the target is not directly affected by
distortion.  But the spatial probability distribution of an immortal
walker is then basically a Gaussian spreading with time, and
relegating sites to a slightly further distance greatly reduces the
chance for a walker starting there to reach the target in a given
number of steps.  Hence rare long and successful walks are
additionally suppressed by shifting the initial distance distribution.

The relative importance of these effects is not easily quantified in
general, as it depends on the absolute lattice size, the regimes, and
the aspect ratio of the lattice, but we will explain their influence
in the different regimes of $L_{1,2}$ and $\ell$.  Any numerical
$\mathcal O(1)$ factors in the comparisons will be omitted.

First, let $\ell\ll L_{1,2}$, the standard large-lattice regime in
which the walk is always of two-dimensional nature.  As argued before,
the different lengths of the lattice hardly have any effect here, and
the encounter probability is governed by the ratio $p\sim
\ell^2/S\ll1$, independent of the aspect ratio.  This region
corresponds to the small peak plateaus at the right of
Figure~\ref{fig:plot-ar-l}.  In this regime, only the static effect
might be important.

For the small-lattice regime $L_{1,2}\ll\ell$, the encounter
probability is close to unity.  A small fraction of walk realizations
does \emph{not} lead to recombination, but in this regime,
\emph{avoiding} the target does not become much easier when the
lattice is distorted.  Now the aspect ratio determines whether the
random walk behaves essentially one- or two-dimensional.  Based on
\eqref{small-lattice-scaling} of Section~\ref{sec:quasi-1d-limit} we
expect the crossover scale between the two types at an aspect ratio of
roughly $\mu=1/\ln L_2$ below which (small-lattice) one-dimensional
behavior prevails, namely $p\approx 1-(L_1/\ell)^2/3$.  The main
difference to the two-dimensional behavior (apart from the absence of
logarithmic corrections) is that only the larger length still enters.
But the respective term has to be small anyway for the expression to
be applicable, and we cannot expect to see it in the plots.  We thus
only have a minute effect of the distortion in this regime as well, a
fact easily seen at the peak plateaus in Figure~\ref{fig:plot-ar-s}.
Actually, the plateaus are more pronounced in this case than for the
corresponding large-lattice plots, because not even the static effect
has any significant influence anymore -- basically the whole lattice
is swept anyway.

Finally consider the intermediate case $L_2\ll\ell\leq L_1$.  This
regime is more complicated since we do not know whether dying or
meeting the target is dominant in setting the residence time of the
walker and the magnitude of $p$.  The aspect ratio determines whether
we deal with a genuinely two-dimensional lattice, or rather with a
lattice so elongated that it is of one-dimensional nature.

If $\mu=L_2/L_1\ll1$ is extremely small, we can effectively consider
the system as one-dimensional and in its 1d-large-lattice regime.
Homogenization in the $L_2$-direction is much faster than the
spreading in the $L_1$-direction then, and thence the probability to
reach the target is essentially that of reaching the projection of the
target position onto the $L_1$-dimension.  This roughly coincides with
the probability for the walker to start within a reach $\ell$ of the
projected position, and this is $p\sim\ell/L_1$.  For the effectively
one-dimensional intermediate lattice this reasoning is justified, as
is shown by Appendix~\ref{app:asy-one-dim}.

In contrast, if the aspect ratio is not small enough for this
viewpoint, the nature of the two-dimensional random walk shows: it
spreads without fully exploring the swept area (rather with increasing
`sponginess' of the set of visited sites, again a specialty of spatial
dimension two).  Consequently, we rather expect an $S/\ell^2$
dependence with typical logarithmic corrections as for the quadratic
case.  $L_2/L_1$ is only moderately small now, so we may use the
result of Section~\ref{sec:montroll-p-star} provided that
$(L_1/\ell)(L_2/\ell)\ll1$.  From this we get
$p\approx1-(S/\ell^2)c_1(\mathrm{const.}+\ln L_1)$, a functional
dependence similar to the small-grain expansion for the quadratic
lattice case.  The crossover aspect ratio between both types of
behavior could not be determined, because in this regime, the
expansion breaks down on the way to the one-dimensional asymptotics.
For the quadratic periodic lattice, the logarithmic correction emerges
from integration of the slowly decaying return probability of long
walks; with a distorted lattice, the dominant bounding contribution
stems from the larger length only, as is most easily seen by rescaling
the approximate integral expression,
Appendix~\ref{app:large-grain-prw}.

This intermediate regime governs the behavior of the
Figures~\ref{fig:plot-ar-s} and~\ref{fig:plot-ar-l} once we leave the
plateau around the quadratic shape.  In all cases, we first enter a
linear decline on the $\log \mu$ scale, as predicted for the
two-dimensional intermediate regime ($\sim -\ln L_1 \propto
\frac12\log\mu$ with $S$ constant).  This behavior roughly starts once
the larger (smaller) length of $L_{1,2}$ becomes of the order of
$\ell$, depending on whether we started in the small-(large-)lattice
regime, which is precisely the condition to enter the intermediate
regime.

For the leftmost (most distorted) part of the plots, the absolute size
of the lattice becomes crucial.  On absolutely large lattices (thick
lines in Figures~\ref{fig:plot-ar-s} and~\ref{fig:plot-ar-l}) the
linear decline ends in an exponential shape close to the fully
distorted lattice.  This is explained by the effective
one-dimensionality as argued above; we end up with the
1d-large-lattice behavior $p\sim\ell/L_1$.  The crossover aspect ratio
is read off to be roughly between $1/400$ and $1/1000$.

In stark contrast, the lattices of small absolute size (thin lines in
Figures~\ref{fig:plot-ar-s} and~\ref{fig:plot-ar-l}) do not show any
clear deviation from the 2d-intermediate-regime decline down to full
distortion.  First, the total decline is not as extreme, as explained
before.  Second, we still are in the 1d-limit for full distortion.
But leaving $\mu=1/S$, immediately two-dimensional effects and the
corresponding logarithmic terms (linear on our scale) dominate the $p$
behavior, because the aspect ratio is far less extreme than for the
absolutely large lattices -- consistently, the crossover aspect ratio
determined there is not reached here.

\section{Conclusions}

In this work, we have thoroughly examined the encounter probability of
two mortal random walkers on a periodic lattice as a model of a
confined geometry.  We compared the results of continuum diffusion
models, exact random walk treatments and their asymptotic behavior and
heuristic results obtained from standard random walk analysis.  We
highlighted their similarities as well as explained the crucial
differences, most importantly the features responsible for the slow
logarithmic convergence to the continuum limit, which is a direct
consequence of the criticality of dimension two for diffusion and
random walks.  The discrete-time random walk results have been shown
to be in excellent agreement with kinetic Monte Carlo simulations of
the continuous-time version that naturally lends itself to physical
applications.  On a side note we could corroborate earlier claims that
the lattice type used is of minor importance to all results discussed
herein.

The second half of this article has been devoted to the analysis of
the influence of the geometry of the lattice, examined at the example
of an aspect ratio differing from unity.  We considered an extremely
distorted lattice and explained our findings for this situation.  Then
we generalized an early result by Montroll to the distorted situation,
and thus could examine the effect of distortion starting from a
quadratic lattice.  Moreover, we determined a scaling limit in which
the dying out of logarithmic terms (characteristic for the
two-dimensional situation) in favor of algebraic corrections (typical
of the one-dimensional case) can be nicely seen and indeed, the
one-dimensional asymptotic behavior is fully recovered.  Finally, we
gave a general explanation of the effects which govern the changing
encounter probability and the regimes to which they pertain.

We believe that this work fills a gap in the vast literature on random
walks and first-passage problems.  Most importantly, we have achieved
the goal stated in the abstract: We have shown that the (completely
analytic) treatment of the simple discrete-time random walk model
provides a full quantitative understanding of the encounter
probability for the (homogeneous) continuous-time reaction-diffusion
system as simulated by the kinetic Monte Carlo method, and this holds
for all the aforementioned aspects including the shape of the lattice
and its structure.  These results are important for many applications
ranging from astrochemistry to biophysical problems, and they equally
matter for both the fundamental theory as well as for the numerical
simulation of reaction-diffusion systems.

Further work will be centered on the role of quenched and annealed
disorder in the rates of hopping and dying of the walkers, mainly from
the point of view of the recombination efficiency of systems such as
those described herein.  An important next step will be the assessment
of the validity of the master equation framework in the homogeneous
situation, where hitherto, spatial correlations between random walkers
on the lattice have been tacitly neglected.

\begin{acknowledgements}
  We thank Ofer Biham for useful discussions.  This work was supported
  by Deutsche Forschungsgemeinschaft within SFB/TR-12
  \textit{Symmetries and Universality in Mesoscopic Systems.}
\end{acknowledgements}

\bibliographystyle{spphys}

\appendix

\section{Derivation of \boldmath{$p_\mathrm{rw}$}}
\label{app:prw}

We closely follow and mimic the notation of \cite{Hughes95} without
citing individual known results.  All of the techniques used here have
been devised quite some time ago, see e.g.\ \cite{Montroll65} with the
exact same formula as \eqref{Pstar}, and we basically put together all
the necessary pieces in an appropriate form.

We deal with a finite periodic homogeneous lattice with
$S=\prod_{j=1}^d L_j$ sites in total, or a $d$-dimensional torus with
extensions $L_j$ in the $j$th direction.  Onto this lattice, we put
two random walkers (in discrete time), starting at random sites,
independently and homogeneously distributed.  They do not interact
except when they meet. The walkers are assumed mortal (corresponding
to desorption) with constant and equal survival probability $\xi$ per
step. Our question is: ``What is the probability that these two
walkers meet before one of them dies?''

This problem can be mapped to that of a single walker, starting from a
random site $s_0$, with the same survival probability per step, the
question being with what probability it eventually reaches a certain
fixed site $s^*$ on the lattice without dying prematurely (all sites are
labeled by a variable $s$ whose structure is irrelevant right now).

Dying and moving of the walker are independent.  Thence
\begin{equation}
\begin{aligned}
 &\Pr\{ \text{Mortal walker reaches site $s^*$ for the}\\
       &\qquad \text{first time on the $k$th step} \} \\
=&\Pr\{ \text{Mortal walker has completed}\\
       &\qquad \text{at least $k$ steps} \} \times\\
&\times \Pr\{ \text{Immortal walker reaches $s^*$ for the}\\
       &\qquad \text{first time on the $k$th step} \} \\
=&\;\xi^k\cdot F_k(s^*|s_0)\,,
\end{aligned}
\end{equation}
where $F_k(s|s_0)$ is the probability (of an immortal random walker)
of arriving at site $s$ \emph{for the first time} on the $k$th step,
given that the walk started at site $s_0$. It should be noted that we
adopt the convention that $F_0(s|s_0)=0$, and therefore do not count a
walker already starting at $s_0$. We do not care for the time
when this first passage of $s^*$ occurs, and are therefore interested
in the quantity
\begin{equation}
F(s^*|s_0;\xi) := \sum_{k=0}^\infty\xi^k F_k(s^*|s_0)\,,
\end{equation}
which happens to be the \emph{generating function} of
$F_k(s^*|s_0)$. One can convince oneself that, by the definition of
$F_k$ as the \emph{first}-passage probability, every encounter is
counted only once, and so $F(s^*|s_0;\xi)$ is the probability of a
mortal random walker with survival rate $\xi$, starting from $s_0$, to
reach $s^*$ before dying.

For the remainder of the derivation we only have to be concerned with
immortal (usual) random walkers. Then for \emph{any} random walk,
there is the relation (e.g.\ \cite{Hughes95} Eq.\ (3.27))
\begin{equation}
\label{Fcentral}
F(s|s_0;\xi) = \frac{P(s|s_0;\xi)-\delta_{s,s_0}}{P(s|s;\xi)}\,,
\end{equation}
where $P(s|s_0;\xi)$ is the generating function of $P_k(s|s_0)$, the
probability that the random walker, starting at site $s_0$, is at site
$s$ on the $k$th step, with the convention that
$P_0(s|s_0)=\delta_{s,s_0}$.

However, regarding the convention used for the first-passage
probability, we want to count a walker that starts at the target site
as ``reaching it for the first time'' on the $0$th step, and not count
later returns. Denoting the corresponding probability by $\tilde
F_k(s|s_0)$, it is related to the original one by
\begin{equation}
\tilde F_k(s|s_0)=\delta_{s,s_0}\delta_{k,0}
+(1-\delta_{s,s_0})F_k(s|s_0)\,.
\end{equation}
For the generating functions this implies
\begin{equation}
\tilde F(s|s_0;\xi)=\delta_{s,s_0}+(1-\delta_{s,s_0})F(s|s_0;\xi)\,,
\end{equation}
and inserting \eqref{Fcentral} yields
\begin{equation}
\label{ourFcentral}
\tilde F(s|s_0;\xi) =\frac{P(s|s_0;\xi)}{P(s|s;\xi)}\,.
\end{equation}

Now we can already write down the answer to our problem in very
general terms. What we actually want is an average $S^{-1}\sum_{s_0
  \in\Omega}\tilde F(s|s_0)$ of the revised first-passage probability
\eqref{ourFcentral} over all starting sites $s_0$ in our finite
periodic lattice $\Omega$, giving us the encounter probability of the
original two random walkers:
\begin{equation}\label{penc}
p_\mathrm{rw} =
\frac{\sum_{s_0\in\Omega}P(s|s_0;\xi)}{S P(s|s;\xi)}\,.
\end{equation}
The apparent dependence on the target site $s$ will vanish in passing
to a homogeneous setting.

For further evaluation, we need to obtain $P(s|s_0;\xi)$. Let us for
the time being denote by this quantity the generating function of the
residence probability of a random walk on an \emph{infinite}
lattice. Furthermore, our walk is homogeneous or translationally
invariant, and we may therefore use a single vector $\vec l$ pointing
from the starting site $s_0$ to the final site $s$ as our variable.
Such `lattice vectors' have $d$ components, all of which take
arbitrary integer values that can be thought of as components of a
`spatial vector' with respect to the $d$ fundamental vectors of the
lattice.  Then it is well-known that (e.g.\ \cite{Hughes95} section
3.3.1)
\begin{equation}
  P(\vec l;\xi)=\frac{1}{(2\pi)^d}\int_{[-\pi,\pi]^d}
  \d^dk\frac{\exp(-i\vec
  l\vec k)}{1-\xi\lambda(\vec k)}\,,
\end{equation}
with the integration domain the first \textsc{Brillouin} zone.
$\lambda(\vec k)$ is the \emph{structure function} of the walk,
defined as
\begin{equation}
\lambda(\vec k)=\sum_{\vec l}\exp(i\vec l\vec k)q(\vec l)\,,
\end{equation}
the sum running over all lattice vectors, and $q(\vec l)$ being the
probability of a step translating by $\vec l$. Since $\sum_{\vec l}
q(\vec l)=1$, $\abs{\lambda(\vec k)}\leq1$.

The transition to a truly finite periodic lattice is now easy:
Attaching a star to the respective residence probabilities, we have
\begin{equation}\label{definePstar}
P_n^*(\vec l)=\sum_{\vec m}P_n(\vec l+\mathbf L\vec m)\,,
\end{equation}
where $\mathbf L=\operatorname{diag}(L_1,\dots,L_d)$.  The sum runs
over all translation vectors of the (infinite) lattice, and this
implies a completely analogous relation for the generating functions.
Obviously, $P^*(\vec l+\mathbf L\vec m)=P^*(\vec l)$ for an arbitrary
$\vec m$ in the infinite lattice. From now on the vector $\vec l$ is
understood to lie in the subset $\Omega$ of the infinite lattice that
stands for the finite periodic lattice (we use the same symbol for
both the finite sets of sites and of associated translation vectors).
It can be shown that consequently we have\footnote{The underlying
  identity reads $\sum_{\vec m}\exp[-i(\mathbf L\vec m)\vec
  k]/(2\pi)^d=\sum_{\vec m}\delta(\vec k-2\pi\mathbf L^{-1}\vec
  m)/(\prod_j L_j)$ and works component-wise.  In a way, this is a
  more general expression than that for the infinite lattice, which
  can be recovered by sending $L_j\to\infty$.}
\begin{equation}
\label{Pstar}
P^*(\vec l;\xi)=\frac{1}{S}\sum_{\vec m\in\Omega}\frac{\exp[-2\pi
  i\vec l(\mathbf L^{-1}\vec m)]}{1-\xi\lambda(2\pi\mathbf L^{-1}\vec m)}\,.
\end{equation}
The sum over the finite lattice $\Omega$ is explicitly a multiple sum
over the components of the vector $\vec m$, each ranging between
$m_j=0\dots L_j-1$.

Rewriting \eqref{penc} for the homogeneous walk and substituting the
appropriate starred probabilities, we obtain
\begin{equation}\label{penc2}
p_\mathrm{rw} =
\frac{\sum_{\vec l\in\Omega}P^*(\vec l;\xi)}{S P^*(\vec 0;\xi)}\,.
\end{equation}
The numerator of this expression is nothing but the $\xi$ transform
of $\sum_{\vec l\in\Omega}P^*_n(\vec l)$, but this is unity due to
conservation of the immortal walker, so that
\begin{equation}
\sum_{\vec l\in\Omega}P^*(\vec l;\xi)=\frac{1}{1-\xi}\,.
\end{equation}
Again using \eqref{Pstar} for the denominator of \eqref{penc2} yields a
general expression for the encounter probability on a homogeneous
regular finite periodic lattice, namely~\eqref{general-prw} of the
main text.

It is often desirable to consider a different convention that does
\emph{not} allow both walkers to start at the same site.  To this end,
one simply has to exclude the lattice distance $\vec 0$ (when talking
about translationally invariant walks) from the average~\eqref{penc},
the result of which is given in the main text as well.  Note that this
is \emph{not} equivalent to simply employing the original
first-passage probability $F$ instead of $\tilde F$, which would allow
this situation, but not appreciate it as an encounter, and would
simply lead to $\xi p_\mathrm{rw}$ instead of $p_\mathrm{rw}$.

\section{Evaluation of one sum}
\label{app:one-sum}
We now restrict ourselves to a still fairly large class of walks,
namely those with structure functions
\begin{multline}
\lambda(\vec k)
=2q_0[\cos k_1+\cos k_2] \\
+2(q_1+q_2)\cos k_1\cos k_2 +2(q_2-q_1)\sin k_1\sin k_2\,.
\end{multline}
Such a structure function belongs to walks with transition
probabilities $q_0$ to take a step into either lattice unit direction,
and $q_1$ and $q_2$ the probabilities to step into direction $(1,1)$
or $(-1,-1)$, and $(1,-1)$ or $(-1,1)$, respectively, subject to the
normalization $4q_0+2(q_1+q_2)=1$.  Clearly, this includes the
aforementioned cases of the main text: The isotropic square lattice
`type (a)' corresponds to $q_0=1/4$ and $q_1=q_2=0$, and the isotropic
triangular lattice walk `type (b)' is represented by $q_0=q_2=1/6$,
$q_1=0$.

For this class of walks, one summation can be explicitly evaluated
\cite{Montroll69}.  The result can be generalized to the $L_1\neq
L_2$ case by some easy accounting work and then reads
\begin{equation}\label{inner-sum}
\begin{split}
P^*(\vec 0;\xi) &= 
\frac{1}{L_1}\sum_{m_1=0}^{L_1-1}[1-2q_0\xi\cos(2\pi m_1/L_1)]^{-1}\\
&\quad \times
[1-\rho_{m_1}^2]^{-1/2}
\frac{1-x_{m_1}^{2L_2}}{1-2x_{m_1}^{L_2}\cos L_2\phi_{m_1}+x_{m_1}^{2L_2}}\,.
\end{split}
\end{equation}
Here, $0<x_{m_1}=[1-(1-\rho_{m_1}^2)^{1/2}]/\rho_{m_1}<1$, and with
\begin{equation}
\begin{aligned}
w_1 &= \frac{2\xi[q_0+(q_1+q_2)\cos(2\pi m_1/L_1)]}
{1-2\xi q_0\cos(2\pi m_1/L_1)},\\
w_2 &= \frac{2\xi(q_1-q_2)\sin(2\pi m_1/L_1)}
{1-2\xi q_0\cos(2\pi m_1/L_1)}\,,
\end{aligned}
\end{equation}
$0<\rho_{m_1}<1$ and $\phi_{m_1}$ are given by $w_1+iw_2
=\rho_{m_1} e^{i\phi_{m_1}}$.  In
general, this yields
\begin{equation}\begin{aligned}
  \rho_{m_1}^2 &= w_1^2+w_2^2 \\
  &=\frac{(2\xi)^2\left[q_0^2+q_1^2+q_2^2+2q_1q_2\cos\frac{4\pi m_1}{L_1}
      +2q_0(q_1+q_2)\cos\frac{2\pi m_1}{L_1}\right]}{
    [1-2\xi q_0\cos(2\pi m_1/L_1)]^2},
\end{aligned}
\end{equation}
and
\begin{equation}
\tan\phi_{m_1}=\frac{(q_1-q_2)\sin\frac{2\pi m_1}{L_1}}{
q_0+(q_1+q_2)\cos\frac{2\pi m_1}{L_1}}
\end{equation}
(whenever well-defined).

For case (a) one obtains $\rho_{m_1}=[2/\xi-\cos(2\pi m_1/L_1)]^{-1}$ and
$\phi_{m_1}=0$, such that
\begin{equation}
P^*(\vec 0;\xi) = 
\frac{1}{L_1}\sum_{m_1=0}^{L_1-1}
\frac{2/\xi}{\sqrt{\rho_{m_1}^{-2}-1}}
 \times\frac{1+x_{m_1}^{L_2}}{1-x_{m_1}^{L_2}}\,.
\end{equation}
For type (b) one has
\begin{equation}
  \rho_{m_1} =\frac{\sqrt{2}\sqrt{1+\cos(2\pi
      m_1/L_1)}}{3/\xi-\cos(2\pi m_1/L_1)}
  =\frac{2\lvert\cos(\pi m_1/L_1)\rvert}{3/\xi-\cos(2\pi m_1/L_1)}
\end{equation}
and
\begin{equation}
  \tan\phi_{m_1}=-\frac{\sin(2\pi m_1/L_1)}{[1+\cos(2\pi m_1/L_1)]}
  =-\tan(\pi m_1/L_1)
\end{equation}
(whenever well-defined), which yields
$\phi_{m_1}=-\pi(m_1/L_1-\left\lfloor 2m_1/L_1\right\rfloor)$.  The
peculiar form of the angle is necessary to assure
$\left\vert\phi_{m_1}\right\vert\leq\pi/2$ corresponding to
$w_1\geq0$, and while the chosen expression may result in a wrong sign
of $\sin\phi_{m_1}$, only the unaffected $\cos(L_2\phi_{m_1})$ is used
in the remainder.  Hence we obtain
\begin{equation}
\begin{split}
P^*(\vec 0;\xi) &= 
\frac{1}{L_1}\sum_{m_1=0}^{L_1-1}\frac{3/\xi}{2\lvert\cos(\pi m_1/L_1)\rvert}
\frac{1}{\sqrt{\rho_{m_1}^{-2}-1}} \\
&\quad \times\frac{1-x_{m_1}^{2L_2}}
{1-2x_{m_1}^{L_2}\cos(L_2\phi_{m_1})+x_{m_1}^{2L_2}}\,,
\end{split}
\end{equation}
and throughout, $x_{m_1}(\rho_{m_1})$ as given above.

The alert reader may object that, in case (b), there is actually one
term for which both $w_1=0=w_2$ and thus $\rho_{m_1}$ vanishes (see
our valid explicit result for the latter), namely for even $L_1$ and
$m_1=L_1/2$.  Hence, $x_{m_1}$, $\phi_{m_1}$ and the last expression
for $P^*(\vec0;\xi)$ are ill-defined then.  From the definition of
$x_{m_1}$ one can see that for $\rho_{m_1}\to0+$,
$x_{m_1}\approx\rho_{m_1}/2\to0+$ as well, and the second line of
\eqref{inner-sum} (corresponding to the evaluated `inner sum' in the
derivation of \cite{Montroll69}) converges to unity, which is the
correct value of the original quantity.  Our last expression complies
with this via canceling singularities -- we chose the simplest form of
the result, which has to be slightly altered for numerical evaluation.

\section{Large lattice approximation}
\label{app:large-grain-prw}
For all approximations in this and the following Section it is
justified to treat $1-\xi$ and $(1-\xi)/\xi=W/a$ synonymously due to
$\xi\lesssim1$, which will no longer be mentioned when it only
introduces higher-order errors compared to the desired accuracy.
Moreover, we will treat logarithms as being of the order of unity.
This might make some expressions more cumbersome, but it is
numerically adequate.

By letting $L_{1,2}\to\infty$ in $P^*(\vec 0;\xi)$
of~\eqref{general-prw} one obtains a double integral, which after two
linear substitutions reads
\begin{equation}
  \frac{p_\mathrm{rw}^{-1}}{(1-\xi)S} = P(\vec 0;\xi) =
  \frac{1}{(2\pi)^2}\int_{[0,2\pi]^2}
  \frac{\d u_1\d u_2}{1-\xi\lambda(\vec u)}\,.
\end{equation}
$\lambda$ periodicity once again allows us to shift the patch of
integration to the \textsc{Brillouin} zone $B$ defined in
Appendix~\ref{app:prw}.  As is shown in the Appendices of
\cite{Hughes95}, in terms of the complete elliptic integral of the
first kind,
\begin{equation}
K(k)=\int_0^1\frac{\d t}{\sqrt{(1-t^2)(1-k^2t^2)}}\,,
\quad\text{with}\quad\abs{k}<1,
\end{equation}
one can derive that $P(\vec 0;\xi)=\frac{2}{\pi} K(\xi)$ for the
square, and
\begin{equation}
P(\vec 0;\xi)=
\frac{6}{\pi\xi\sqrt{(c_-+1)(c_+-1)}}
\times\,K\left(\sqrt{\frac{2(c_+-c_-)}{(c_-+1)(c_+-1)}}\right)
\end{equation}
for the triangular lattice, where $c_\pm=3/\xi+1\pm\sqrt{3+6/\xi}$.
Now we need the expansion of the elliptic integral for $k\lesssim1$
(which is also the case for the triangular lattice if $\xi$ itself is
close to unity), viz.\
\begin{multline}
K(k)=\sum_{n=0}^{\infty}\left[\frac{\left(\frac12\right)_n}{n!}\right]^2
(1-k^2)^n\\
\times\,\left[-(1/2)\ln(1-k^2)+\psi(n+1)-\psi(n+1/2)\right],
\end{multline}
here $(a)_n=\Gamma(n+a)/\Gamma(a)$ is the \textsc{Pochhammer} symbol,
equal to $a(a+1)\dots(a+n-1)$ for positive integer $n$.  Using these
expressions in the exact results for $P(\vec 0;\xi)$, one obtains
\begin{equation}
P(\vec 0;\xi)=[1+\order(1-\xi)]
\times\,\begin{cases}
\frac1\pi\ln[8/(1-\xi)] & \text{square lattice,}\\
\frac{\sqrt{3}}{2\pi}\ln[12/(1-\xi)] & \text{triangular lattice.}
\end{cases}
\end{equation}
This yields the expressions for $p_\mathrm{rw}$ presented in the main
text.

\section{Simulation results for $p_\mathrm{mc}$}
\label{app:sims-pmc-vs-prw}

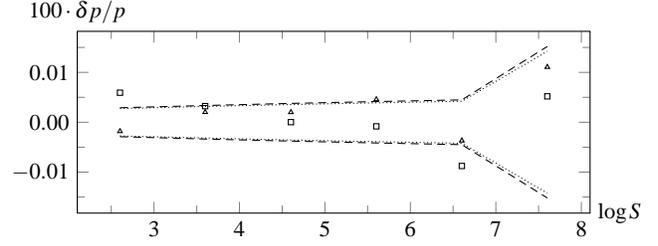
\begin{figure}
  \beginpgfgraphicnamed{graphic3a}
  \begin{tikzpicture}
    \begin{axis}[xlabel=$\log S$,ylabel=$100\cdot\delta p/p$,
      y tick label style={/pgf/number format/precision=2},
      width=\columnwidth,height=40mm,try min ticks=5]
      \addplot[densely dashed] plot file {dat/mcvsrw_sq-4.s.dat};
      \addplot[densely dotted] plot file {dat/mcvsrw_tr-4.s.dat};
      \pgfplotsset{/pgfplots/y filter/.code={\pgfmathmultiply{#1}{-1}}}
      \addplot[densely dashed] plot file {dat/mcvsrw_sq-4.s.dat};
      \addplot[densely dotted] plot file {dat/mcvsrw_tr-4.s.dat};
      \pgfplotsset{/pgfplots/y filter/.code={}}
      \addplot[only marks,mark=square] plot file {dat/mcvsrw_sq-4.dat};
      \addplot[only marks,mark=triangle] plot file {dat/mcvsrw_tr-4.dat};
    \end{axis}
  \end{tikzpicture}
  \endpgfgraphicnamed
  \caption{Relative difference of $\tilde p_\mathrm{mc}$
    w.r.t.\ $\tilde p_\mathrm{rw}$ for $SW/(4a)=10^{-4}$, type (a)
    (dashed\,/\,squares), type (b) (dotted\,/\,triangles).}
  \label{fig:delta-4}
\end{figure}

\begin{figure}
  \beginpgfgraphicnamed{graphic3b}
  \begin{tikzpicture}
    \begin{axis}[xlabel=$\log S$,ylabel=$100\cdot\delta p/p$,
      y tick label style={/pgf/number format/precision=2},
      width=\columnwidth,height=40mm]
      \addplot[densely dashed] plot file {dat/mcvsrw_sq-3.s.dat};
      \addplot[densely dotted] plot file {dat/mcvsrw_tr-3.s.dat};
      \pgfplotsset{/pgfplots/y filter/.code={\pgfmathmultiply{#1}{-1}}}
      \addplot[densely dashed] plot file {dat/mcvsrw_sq-3.s.dat};
      \addplot[densely dotted] plot file {dat/mcvsrw_tr-3.s.dat};
      \pgfplotsset{/pgfplots/y filter/.code={}}
      \addplot[only marks,mark=square] plot file {dat/mcvsrw_sq-3.dat};
      \addplot[only marks,mark=triangle] plot file {dat/mcvsrw_tr-3.dat};
    \end{axis}
  \end{tikzpicture}
  \endpgfgraphicnamed
  \caption{$SW/(4a)=10^{-3}$.}\label{fig:delta-3}
\end{figure}
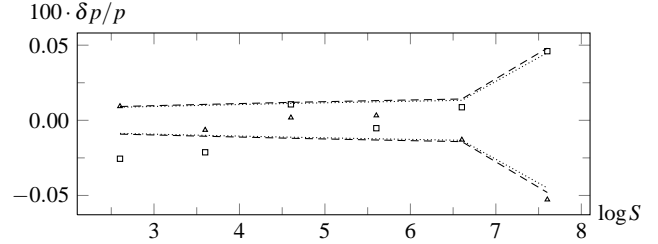

\begin{figure}
  \beginpgfgraphicnamed{graphic3c}
  \begin{tikzpicture}
    \begin{axis}[xlabel=$\log S$,ylabel=$100\cdot\delta p/p$,
      y tick label style={/pgf/number format/precision=1},
      width=\columnwidth,height=40mm,try min ticks=5]
      \addplot[densely dashed] plot file {dat/mcvsrw_sq-2.s.dat};
      \addplot[densely dotted] plot file {dat/mcvsrw_tr-2.s.dat};
      \pgfplotsset{/pgfplots/y filter/.code={\pgfmathmultiply{#1}{-1}}}
      \addplot[densely dashed] plot file {dat/mcvsrw_sq-2.s.dat};
      \addplot[densely dotted] plot file {dat/mcvsrw_tr-2.s.dat};
      \pgfplotsset{/pgfplots/y filter/.code={}}
      \addplot[only marks,mark=square] plot file {dat/mcvsrw_sq-2.dat};
      \addplot[only marks,mark=triangle] plot file {dat/mcvsrw_tr-2.dat};
    \end{axis}
  \end{tikzpicture}
  \endpgfgraphicnamed
  \caption{$SW/(4a)=10^{-2}$.}\label{fig:delta-2}
\end{figure}
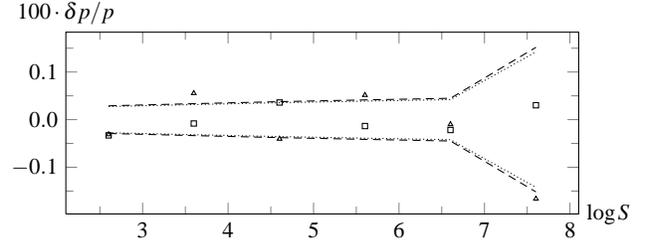

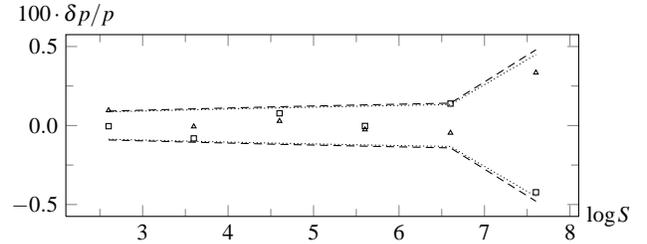
\begin{figure}
  \beginpgfgraphicnamed{graphic3d}
  \begin{tikzpicture}
    \begin{axis}[xlabel=$\log S$,ylabel=$100\cdot\delta p/p$,
      y tick label style={/pgf/number format/precision=1},
      width=\columnwidth,height=40mm]
      \addplot[densely dashed] plot file {dat/mcvsrw_sq-1.s.dat};
      \addplot[densely dotted] plot file {dat/mcvsrw_tr-1.s.dat};
      \pgfplotsset{/pgfplots/y filter/.code={\pgfmathmultiply{#1}{-1}}}
      \addplot[densely dashed] plot file {dat/mcvsrw_sq-1.s.dat};
      \addplot[densely dotted] plot file {dat/mcvsrw_tr-1.s.dat};
      \pgfplotsset{/pgfplots/y filter/.code={}}
      \addplot[only marks,mark=square] plot file {dat/mcvsrw_sq-1.dat};
      \addplot[only marks,mark=triangle] plot file {dat/mcvsrw_tr-1.dat};
    \end{axis}
  \end{tikzpicture}
  \endpgfgraphicnamed
  \caption{$SW/(4a)=10^{-1}$.}\label{fig:delta-1}
\end{figure}

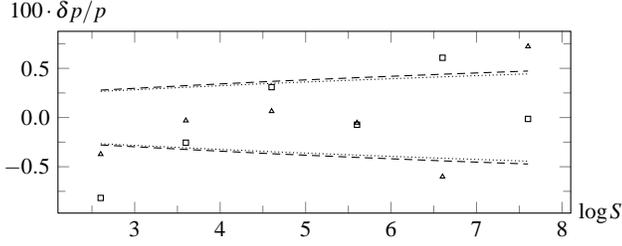
\begin{figure}
  \beginpgfgraphicnamed{graphic3e}
  \begin{tikzpicture}
    \begin{axis}[xlabel=$\log S$,ylabel=$100\cdot\delta p/p$,
      y tick label style={/pgf/number format/precision=1},
      width=\columnwidth,height=40mm,try min ticks=5]
      \addplot[densely dashed] plot file {dat/mcvsrw_sq0.s.dat};
      \addplot[densely dotted] plot file {dat/mcvsrw_tr0.s.dat};
      \pgfplotsset{/pgfplots/y filter/.code={\pgfmathmultiply{#1}{-1}}}
      \addplot[densely dashed] plot file {dat/mcvsrw_sq0.s.dat};
      \addplot[densely dotted] plot file {dat/mcvsrw_tr0.s.dat};
      \pgfplotsset{/pgfplots/y filter/.code={}}
      \addplot[only marks,mark=square] plot file {dat/mcvsrw_sq0.dat};
      \addplot[only marks,mark=triangle] plot file {dat/mcvsrw_tr0.dat};
    \end{axis}
  \end{tikzpicture}
  \endpgfgraphicnamed
  \caption{$SW/(4a)=10^0$.}\label{fig:delta0}
\end{figure}

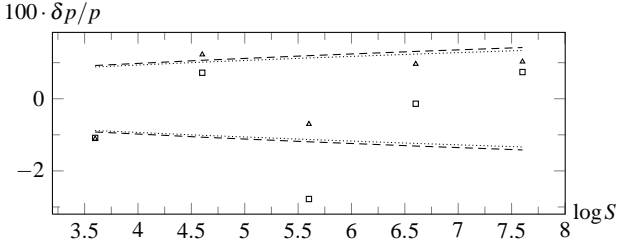
\begin{figure}
  \beginpgfgraphicnamed{graphic3f}
  \begin{tikzpicture}
    \begin{axis}[xlabel=$\log S$,ylabel=$100\cdot\delta p/p$,
      y tick label style={/pgf/number format/precision=0},
      width=\columnwidth,height=40mm]
      \addplot[densely dashed] plot file {dat/mcvsrw_sq1.s.dat};
      \addplot[densely dotted] plot file {dat/mcvsrw_tr1.s.dat};
      \pgfplotsset{/pgfplots/y filter/.code={\pgfmathmultiply{#1}{-1}}}
      \addplot[densely dashed] plot file {dat/mcvsrw_sq1.s.dat};
      \addplot[densely dotted] plot file {dat/mcvsrw_tr1.s.dat};
      \pgfplotsset{/pgfplots/y filter/.code={}}
      \addplot[only marks,mark=square] plot file {dat/mcvsrw_sq1.dat};
      \addplot[only marks,mark=triangle] plot file {dat/mcvsrw_tr1.dat};
    \end{axis}
  \end{tikzpicture}
  \endpgfgraphicnamed
  \caption{$SW/(4a)=10^1$.}\label{fig:delta1}
\end{figure}

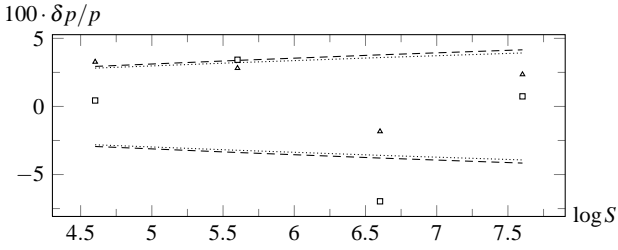
\begin{figure}
  \beginpgfgraphicnamed{graphic3g}
  \begin{tikzpicture}
    \begin{axis}[xlabel=$\log S$,ylabel=$100\cdot\delta p/p$,
      y tick label style={/pgf/number format/precision=0},
      width=\columnwidth,height=40mm]
      \addplot[densely dashed] plot file {dat/mcvsrw_sq2.s.dat};
      \addplot[densely dotted] plot file {dat/mcvsrw_tr2.s.dat};
      \pgfplotsset{/pgfplots/y filter/.code={\pgfmathmultiply{#1}{-1}}}
      \addplot[densely dashed] plot file {dat/mcvsrw_sq2.s.dat};
      \addplot[densely dotted] plot file {dat/mcvsrw_tr2.s.dat};
      \pgfplotsset{/pgfplots/y filter/.code={}}
      \addplot[only marks,mark=square] plot file {dat/mcvsrw_sq2.dat};
      \addplot[only marks,mark=triangle] plot file {dat/mcvsrw_tr2.dat};
    \end{axis}
  \end{tikzpicture}
  \endpgfgraphicnamed
  \caption{$SW/(4a)=10^2$.}\label{fig:delta2}
\end{figure}

Figures~\ref{fig:delta-4}--\ref{fig:delta2} show the relative error
(per cent) $100\cdot(\tilde p_\mathrm{mc}-\tilde p_\mathrm{rw})/\tilde
p_\mathrm{rw}$ as a function of the lattice size $S$, for quadratic
lattices.  Lattice sizes and rate ratios $a/W$ are chosen as described
in Section~\ref{sec:comparison-results}, so that one plot roughly
corresponds to a fixed regime and about constant $p$, to keep the
standard deviation of the same order of magnitude.  Outside these
parameter ranges, nothing interesting happens; in the large lattice
regime, the leftmost data points are omitted as they no longer satisfy
$a/W\gg1$.  We have plotted a corridor of half width $\tilde\sigma$
around perfect coincidence to show that the discrepancy between
analysis and simulations is statistically insignificant.
\noindent
\noindent
\noindent
\noindent
\noindent
\noindent

Figures~\ref{fig:ar-2l}--\ref{fig:ar-6s} show corresponding simulation
results for rectangular lattices of varying aspect ratio.  Here $S$ is
constant for one plot.  We restrict ourselves to $S=400$ and
$S=4\times10^6$ as examples of absolutely small and large lattices,
respectively.  Further, we keep the ``regime'' (in the original sense
introduced for quadratic lattices) constant as well: for $S=400$, we
choose $a/W=100$ and $a/W=10^4$, for $S=4\times10^6$ we test with
$a/W=10^6$ and $a/W=10^8$ (the latter figures belonging to the
small-lattice regime), but note that the refined regimes for $L_1\neq
L_2$ actually change when distorting the lattice.  We have chosen our
values so that the larger length $L_1$ exceeds the random walk length
$\ell $ in the large-lattice regime throughout, and that it is much
smaller than this length in the small-lattice regime for the quadratic
case, but finally becoming much larger than it when distorting the
lattice.  Again, we show the relative error of simulations with
respect to the random walk result in per cent, and plot a standard
deviation (as obtained from $\tilde p_ \mathrm{rw}$) corridor for
comparison.
\begin{figure}
  \beginpgfgraphicnamed{graphic4a}
  \begin{tikzpicture}
    \begin{axis}[xlabel=$\log L_2$,ylabel=$100\cdot\delta p/p$,
      y tick label style={/pgf/number format/precision=1},
      width=\columnwidth,height=40mm]
      \addplot[densely dashed] plot file {dat/mcvsrwar_sq2l.s.dat};
      \addplot[densely dotted] plot file {dat/mcvsrwar_tr2l.s.dat};
      \pgfplotsset{/pgfplots/y filter/.code={\pgfmathmultiply{#1}{-1}}}
      \addplot[densely dashed] plot file {dat/mcvsrwar_sq2l.s.dat};
      \addplot[densely dotted] plot file {dat/mcvsrwar_tr2l.s.dat};
      \pgfplotsset{/pgfplots/y filter/.code={}}
      \addplot[only marks,mark=square] plot file {dat/mcvsrwar_sq2l.dat};
      \addplot[only marks,mark=triangle] plot file {dat/mcvsrwar_tr2l.dat};
    \end{axis}
  \end{tikzpicture}
  \endpgfgraphicnamed
  \caption{Relative difference of $\tilde p_\mathrm{mc}$ w.r.t.\
    $\tilde p_\mathrm{rw}$ for $S=4\times10^2$, $W/a=10^{-2}$, type (a)
    (dashed\,/\,squares), type (b) (dotted\,/\,triangles).}
  \label{fig:ar-2l}
\end{figure}
\begin{figure}
  \beginpgfgraphicnamed{graphic4b}
  \begin{tikzpicture}
    \begin{axis}[xlabel=$\log L_2$,ylabel=$100\cdot\delta p/p$,
      y tick label style={/pgf/number format/precision=1},
      width=\columnwidth,height=40mm]
      \addplot[densely dashed] plot file {dat/mcvsrwar_sq2s.s.dat};
      \addplot[densely dotted] plot file {dat/mcvsrwar_tr2s.s.dat};
      \pgfplotsset{/pgfplots/y filter/.code={\pgfmathmultiply{#1}{-1}}}
      \addplot[densely dashed] plot file {dat/mcvsrwar_sq2s.s.dat};
      \addplot[densely dotted] plot file {dat/mcvsrwar_tr2s.s.dat};
      \pgfplotsset{/pgfplots/y filter/.code={}}
      \addplot[only marks,mark=square] plot file {dat/mcvsrwar_sq2s.dat};
      \addplot[only marks,mark=triangle] plot file {dat/mcvsrwar_tr2s.dat};
    \end{axis}
  \end{tikzpicture}
  \endpgfgraphicnamed
  \caption{$S=4\times10^2$, $W/a=10^{-4}$.}\label{fig:ar-2s}
\end{figure}
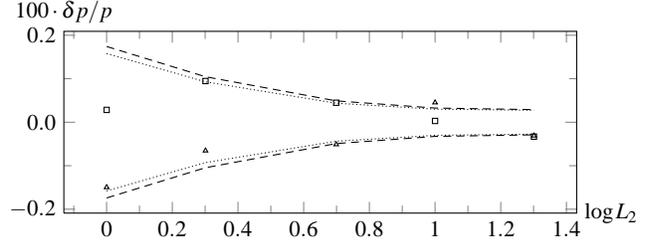
\begin{figure}
  \beginpgfgraphicnamed{graphic4c}
  \begin{tikzpicture}
    \begin{axis}[xlabel=$\log L_2$,ylabel=$100\cdot\delta p/p$,
      y tick label style={/pgf/number format/precision=0},
      width=\columnwidth,height=40mm]
      \addplot[densely dashed] plot file {dat/mcvsrwar_sq6l.s.dat};
      \addplot[densely dotted] plot file {dat/mcvsrwar_tr6l.s.dat};
      \pgfplotsset{/pgfplots/y filter/.code={\pgfmathmultiply{#1}{-1}}}
      \addplot[densely dashed] plot file {dat/mcvsrwar_sq6l.s.dat};
      \addplot[densely dotted] plot file {dat/mcvsrwar_tr6l.s.dat};
      \pgfplotsset{/pgfplots/y filter/.code={}}
      \addplot[only marks,mark=square] plot file {dat/mcvsrwar_sq6l.dat};
      \addplot[only marks,mark=triangle] plot file {dat/mcvsrwar_tr6l.dat};
    \end{axis}
  \end{tikzpicture}
  \endpgfgraphicnamed
  \caption{$S=4\times10^6$, $W/a=10^{-6}$.}\label{fig:ar-6l}
\end{figure}
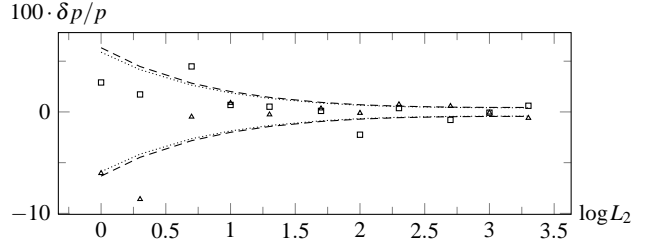
\begin{figure}
  \beginpgfgraphicnamed{graphic4d}
  \begin{tikzpicture}
    \begin{axis}[xlabel=$\log L_2$,ylabel=$100\cdot\delta p/p$,
      y tick label style={/pgf/number format/precision=0},
      width=\columnwidth,height=40mm]
      \addplot[densely dashed] plot file {dat/mcvsrwar_sq6s.s.dat};
      \addplot[densely dotted] plot file {dat/mcvsrwar_tr6s.s.dat};
      \pgfplotsset{/pgfplots/y filter/.code={\pgfmathmultiply{#1}{-1}}}
      \addplot[densely dashed] plot file {dat/mcvsrwar_sq6s.s.dat};
      \addplot[densely dotted] plot file {dat/mcvsrwar_tr6s.s.dat};
      \pgfplotsset{/pgfplots/y filter/.code={}}
      \addplot[only marks,mark=square] plot file {dat/mcvsrwar_sq6s.dat};
      \addplot[only marks,mark=triangle] plot file {dat/mcvsrwar_tr6s.dat};
    \end{axis}
  \end{tikzpicture}
  \endpgfgraphicnamed
  \caption{$S=4\times10^6$, $W/a=10^{-8}$.}\label{fig:ar-6s}
\end{figure}
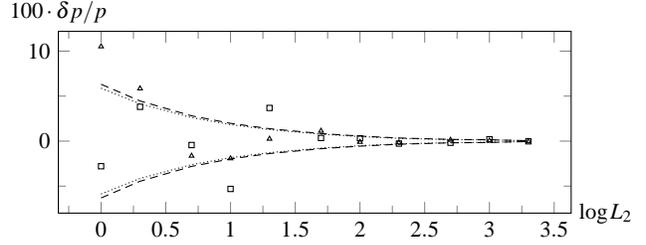

\section{Asymptotics of the truly one-dimensional result}
\label{app:asy-one-dim}

For the one-dimensional symmetric random walk with structure function
$\lambda(k)=\cos k$ and a lattice of $S\equiv L$ sites, we have
\begin{equation}
  P^*(0;\xi)=\frac1L \sum_{m=0}^{L-1}\frac1{1-\xi\cos(2\pi m/L)}\,.
\end{equation}
The identity of Appendix A in \cite{Montroll69} (with $w_1=\xi$,
$w_2=0$, so that $\rho_{m_1}=\xi$ and $\phi_{m_1}\equiv0$, and
$x_{m_1}=[1-(1-\xi^2)^{1/2}]/\xi$) then yields
\begin{equation}
    P^*(0;\xi) = \frac{1}{\sqrt{1-\xi^2}}\frac{1+x_{m_1}^L}{1-x_{m_1}^L}
    = \frac{1}{\sqrt{1-\xi^2}}
    \frac{\xi^L+[1-(1-\xi^2)^{1/2}]^L}{\xi^L-[1-(1-\xi^2)^{1/2}]^L}\,.
\end{equation}

For comparison with the asymptotic two-dimensional behavior, we
evaluate the asymptotics of the 1d-expression for a `small' lattice,
i.e.\ for $L\ll \ell =\sqrt{\xi/(1-\xi)}$.  Let
$\alpha=\sqrt{1-\xi^2}\ll1$, and note that this is \emph{not} exactly
the $\alpha$ of \cite{Montroll69}.  However, $L/\ell \approx
L\sqrt{1-\xi}= L\sqrt{1-\sqrt{1-\alpha^2}}\approx L\alpha/\sqrt{2}$,
and since this is to be $\ll1$, we can still use $L\alpha\ll1$ in the
following.  With $1-\xi=1-\sqrt{1-\alpha^2}$ and
$p_\mathrm{rw}^{-1}=L(1-\xi)P^*(0;\xi)$ we then obtain after minor
manipulations
\begin{equation}
  \label{p1d-still-ex}
  p_\mathrm{rw,\,1d}^{-1}= \frac{L[1-(1-\alpha^2)^{1/2}]}{\alpha}
  \frac{1+\left(\frac{1-\alpha}{1+\alpha}\right)^{L/2}}
  {1-\left(\frac{1-\alpha}{1+\alpha}\right)^{L/2}}\,.
\end{equation}
Straight-forward expansion in $L\alpha\ll1$ yields
\begin{equation}\label{p1d-small-lattice}
  p_\mathrm{rw,\,1d}^{-1}= 1+\frac{(L\alpha)^2}{12}+\order(L\alpha)^3\,,
\end{equation}
where we omitted all terms of relative order $\alpha$.  This is
sufficient for the sought scaling limit $L\to\infty$, $\alpha\to0$
while $L\alpha\simeq\const$, and it is additionally justified by a
\textsc{Mathematica} check of our calculations, which shows that the
only interesting $\alpha$ orders left out in the result are terms of
$\order(\alpha^2)$ and thus much smaller than
$(L\alpha)^2=L^2(1-\xi^2)\approx2(L/\ell )^2$.

Obviously, the most important difference compared to the 2d case is
the absence of logarithmic terms, which are a characteristic sign of
the marginal dimension two.

For completeness, let us also give the large-lattice asymptotics,
i.e., for the case $L\alpha\gg1$.  Starting from the still
exact~\eqref{p1d-still-ex}, we use
$(1-\alpha)/(1+\alpha)=1-2\alpha+\mathcal{O}(\alpha^2)$.  For
$L\alpha\gg1$, this raised to the $L/2$ power is much smaller than
unity, the correct expansion thus reading
$\exp(-L\alpha)(1+\mathcal{O}(L\alpha^2))$, and here we assume that
terms of the latter relative order can be omitted as being much
smaller than unity -- this is an example that may be refined as
necessary.  One thus obtains the large-lattice result
\begin{equation}
  \begin{split}
  p_\mathrm{rw,\,1d} &= \frac{2}{L\alpha}\left(1-2\e^{-L\alpha}
    +\mathcal{O}(\e^{-L\alpha}L\alpha^2,\, \e^{-2L\alpha})\right)\\
  &\approx \sqrt{2}\frac{\ell }{L}
  \left(1-2\e^{-\sqrt{2}L/\ell }\right)\,.
\end{split}
\end{equation}
Note that in this regime, the ratio of lengths no longer appears
squared.  In one dimension and for large lattices, the essential
question for the encounter probability is whether the walker is
deposited in the range $\ell $ from the target, and the probability
for this to happen is the length ratio raised to the lattice
dimension.  The difference to the two-dimensional case is that the 1d
random walk explores a dense region instead of a sponge-like
structure.

Lastly, corresponding asymptotics for the extremely distorted version
of the originally two-dimensional random walk are obtained by a mere
rescaling $\ell \to\ell /\sqrt{2}$, as can be
seen from Section~\ref{sec:extreme-dist}.

\end{document}